\documentclass[10pt,a4paper]{article}
 
\usepackage[export]{adjustbox}
\usepackage{stmaryrd}
 
\usepackage{graphics,graphicx}
\usepackage{appendix}
 
\usepackage{amsmath,amssymb,dsfont,amsthm}
\RequirePackage{amsthm}
\usepackage{caption}
 
\usepackage{subfig}
 
\if@onethmnum
  \newtheorem{theorem}{Theorem}
  \newtheorem{lemma}[theorem]{Lemma}
  \newtheorem{corollary}[theorem]{Corollary}
  \newtheorem{proposition}[theorem]{Proposition}
  \newtheorem{definition}[theorem]{Definition}
\else
  \newtheorem{theorem}{Theorem}[section]
  \newtheorem{lemma}[theorem]{Lemma}
  
  \newtheorem{proposition}[theorem]{Proposition}
  \newtheorem{definition}[theorem]{Definition}
\fi

\newtheorem{algorithm}{Algorithm}
\DeclareMathOperator*{\argmin}{arg\,min}

\title{\uppercase{Universal Prediction Distribution for Surrogate Models}}
 
\author{Malek Ben Salem\footnotemark[1] \footnotemark[2]
\and Olivier Roustant\footnotemark[1]
\and Fabrice Gamboa \footnotemark[3]
\and Lionel Tomaso \footnotemark[2]}

\begin{document}
\maketitle
 
\newcommand{\norme}[1]{\left\Vert #1\right\Vert}
\renewcommand{\thefootnote}{\fnsymbol{footnote}}
\footnotetext[1]{EMSE  Ecole des Mines de St-Etienne, UMR CNRS 6158, LIMOS, F-42023: 158 Cours Fauriel, Saint-Etienne.}
\footnotetext[2]{ANSYS, Inc: 11 Avenue Albert Einstein F-69100 Villeurbanne. }
\footnotetext[3]{IMT Institut de Mathématiques de Toulouse: 118 route de Narbonne,  31062 TOULOUSE Cedex 9.}
%\footnotetext{\email{malek.ben-salem@emse.fr}}
\renewcommand{\thefootnote}{\arabic{footnote}}

\begin{abstract}
The use of surrogate models instead of computationally expensive simulation codes is very convenient in engineering. Roughly speaking, there are two kinds of surrogate models: the deterministic and the  probabilistic ones. These last are generally based on Gaussian assumptions.  The main advantage of probabilistic approach is that it provides a measure of uncertainty associated with the surrogate model in the whole space. This uncertainty is an efficient tool to construct strategies for various problems such as prediction enhancement, optimization or inversion.

In this paper, we propose a universal method to define a measure  of  uncertainty suitable for any surrogate model either deterministic or probabilistic. It relies on Cross-Validation (CV) sub-models predictions. This empirical distribution may be computed in much more general frames than the Gaussian one. So that it is called the Universal Prediction distribution (\textit{UP distribution}).
It allows  the definition of many sampling criteria. We give and study  adaptive sampling techniques for global refinement and an extension of the  so-called Efficient Global Optimization (EGO) algorithm.  We also discuss  the use of the \textit{UP distribution}  for inversion problems. The performances of these new algorithms are studied both on toys models and on an engineering design problem.
\end{abstract}

\textbf{keywords} Surrogate models, Design of experiments, Bayesian optimization 

%\begin{AMS} 62L05,  62K05, 90C26 \end{AMS}

%\pagestyle{myheadings}
%\thispagestyle{plain}
%\markboth{BEN SALEM, ROUSTANT, GAMBOA and TOMASO}{Surrogate Models Universal Prediction Distribution}

\section{Introduction}
 Surrogate modeling techniques are widely used and studied in engineering and research. Their main purpose is to replace an expensive-to-evaluate  function $s$ by a simple response surface $ \hat{s}$ also called   surrogate model or  meta-model. Notice that $s$ can be a  computation-intensive  simulation code. These  surrogate models are  based on a given training set of $n$ observations $z_j=(x_j,y_j)$ where $1\leq j\leq n$ and $ y_j = s(x_j)$. The accuracy of the surrogate model relies, \textit{inter alia}, on the relevance of the training set. The aim of surrogate modeling  is generally to estimate some features of the function $s$  using $ \hat{s}$. Of course one is looking for the best trade-off between a good accuracy of the feature estimation and the number of  calls of $s$. Consequently, the design of experiments (DOE), that is the sampling of $(x_j)_{1\leq j\leq n}$, is a crucial step and  an active research field.
 
 There are two ways to sample: either drawing the training set $(x_j)_{1\leq j\leq n}$ at once or  building it sequentially.  Among the sequential techniques, some  are based  on surrogate models. They  rely on the feature of $s$ that one wishes to estimate. Popular examples are the EGO \cite{jonesEGO} and the Stepwise Uncertainty Reduction  (SUR) \cite{bect2012}. These two methods  use Gaussian process  regression  also called kriging model. It is a widely used surrogate modeling technique. Its popularity is mainly due  to its statistical nature and properties.  Indeed, it is a Bayesian  inference  technique for functions. In this stochastic frame, it provides an estimate of the prediction error distribution.  This distribution is the main tool in Gaussian surrogate sequential designs.  For instance, it allows  the introduction and the computation of different sampling criteria such as the Expected Improvement (EI) \cite{jonesEGO} or the Expected Feasibility (EF) \cite{bichon2008efficient}.
Away from the Gaussian case, many  surrogate models are also available and useful. Notice that none of them including the Gaussian process surrogate model are the best in all circumstances \cite{Sumoart11}. 
Classical surrogate models are for instance support vector machine \cite{svm}, linear regression \cite{qr1}, moving least squares  \cite{mls}. More recently a mixture of surrogates has been considered in \cite{viana2009multiple,goelensemble}. Nevertheless, these methods are generally not naturally embeddable in some stochastic frame. Hence, they do not provide any prediction error distribution. To overcome this drawback, several empirical design techniques have been discussed in the literature. These techniques are generally based on resampling methods such as bootstrap, jackknife, or cross-validation. For instance, Gazut et al. \cite{Gazut2008} and Jin et al. \cite{jin2002sequential} consider a population of surrogate models constructed by resampling the available data using bootstrap or cross-validation. Then, they compute  the empirical  variance of the predictions of these surrogate models. Finally, they sample iteratively the point that maximizes the empirical variance in order to improve the accuracy of the prediction. To perform optimization, Kleijnen et al. \cite{kleijnen2012expected}  use a bootstrapped kriging variance instead of the kriging variance to compute the expected improvement. Their algorithm  consists in maximizing the expected improvement  computed through bootstrapped kriging variance. However, most of these resampling method-based design techniques lead to clustered designs \cite{aut,jin2002sequential}.

In this paper, we give a general way to build an empirical prediction  distribution allowing sequential design strategies in a very broad frame.  
Its support is the set of all the predictions obtained by the cross-validation surrogate models. The novelty of our approach  is that it provides a  prediction uncertainty distribution. This allows  a large set of sampling criteria. Furthermore, it leads naturally to non-clustered designs as explained  in Section \ref{sec:Ref}. 

The paper is organized as follows. We start by  presenting in  Section \ref{sec:notations} the background and notations.
In Section \ref{sec:up} we introduce the Universal Prediction (UP) empirical distribution. In Sections  \ref{sec:Ref} and \ref{sec:Opt}, we use and study features estimation and the corresponding sampling schemes  built on the UP empirical distribution.  Section \ref{sec:Ref} is devoted to  the enhancement of the overall model accuracy. Section \ref{sec:Opt}  concerns optimization.  In Section \ref{sec:mix}, we study a real life industrial case  implementing the methodology developed in Section \ref{sec:Ref}. Section \ref{sec:empinversion} deals with the inversion problem. 
In Section \ref{sec:conclusion}, we conclude and discuss  the possible extensions of our work. All proofs are postponed to Section \ref{sec:proofs}.

\section{Background and notations} \label{sec:notations}
 
\subsection{General notation}\label{sec:form}

 To begin with, let $s$ denote a real-valued function defined on $\mathbb{X}$, a nonempty compact subset of the Euclidean space $\mathbb{R}^p$ ($p \in \mathbb{N}^\star$). In order to estimate $s$, we have at hand a sample of size $n$ ($n \geq 2$):
 $\mathbf{X_n} = \begin{pmatrix} \mathbf{x_1}, &  \dots, & \mathbf{x_n} \end{pmatrix}^\top$ with $\mathbf{x_j} \in \mathbb{X}$, $j \in \llbracket 1; n \rrbracket $ and $\mathbf{Y_n}= \begin{pmatrix} y_1,&  \dots, & y_n \end{pmatrix}^\top$ where $y_j = s (\mathbf{x_j})$ for  $j\in \llbracket 1; n \rrbracket $. We note $\mathbf{Y_n} = s(\mathbf{X_n})$.
 
 Let $\mathbf{Z_n}$ denote the observations: $\mathbf{Z_n}:= \{ (\mathbf{x_j},y_j),  j\in \llbracket 1; n \rrbracket \}$.  Using $\mathbf{Z_n}$, we build a surrogate model $ \hat{s}_{n}$ that mimics the behaviour of $s$. For example, $ \hat{s}_{n}$ can be a second order polynomial regression model. %Let $\mathbb{S}$  
For $i \in \{ 1\hdots n  \}$, we set $\mathbf{Z}_{n,-i}:= \{(\mathbf{x_j},y_j),  j=1,\hdots, n,  j \neq i  \}$  and so $ \hat{s}_{n,-i}$ is the surrogate model obtained by using only the dataset $\mathbf{Z}_{n,-i}$. We will call $ \hat{s}_{n}$ the master surrogate model and $(\hat{s}_{n,-i})_{i=1\hdots n}$ its sub-models. 

 Further, let $d(.,.)$ denote a given distance on $\mathbb{R}^p$ (typically the Euclidean one). For $\mathbf{x} \in \mathbb{X}$ and $A \subset \mathbb{X}$, we set:
 $\underline{d}_A (\mathbf{x}) = \inf \{ d(\mathbf{x},\mathbf{x'}): \mathbf{x'} \in A  \} $ and if $A=\{\mathbf{x'_1}, \hdots,\mathbf{x'_m}\}$ is finite ($m\in \mathbb{N}^\star$), for $i \in 1,\dots,m$ let $A_{-i}$ denote $\{\mathbf{x'_j}, j=1\hdots m, j \neq i\}$. Finally,  we set $\bar{d}(A) = \max \{ \underline{d}_{A_{-i}} (\mathbf{x'_i}): i =1,\hdots,m  \}$, the largest distance of an element of $A$ to its nearest neighbor.

\subsection{Cross-validation}
Training an algorithm and evaluating its statistical performances on the same data yields an optimistic result \cite{Arl_Cel2010}. It is well known that it is easy to over-fit the data by including too many degrees of freedom and so inflate  the fit statistics.
The idea behind Cross-validation (CV)  is to estimate the risk of an algorithm splitting the dataset once or several times. One part of  the data (the training set) is used for training and the remaining one (the validation set) is used for estimating the risk of the algorithm. 
Simple validation or hold-out \cite{Devroye79} is hence a cross-validation technique. It relies on one splitting of the data. Then  one set is used as training set and the second one is used as validation set. Some other CV techniques consist in a repetitive generation of hold-out estimator with different data splitting \cite{geisser1975predictive}. One can cite, for instance, the Leave-One-Out Cross-Validation (LOO-CV) and the $K$-Fold Cross-Validation (KFCV).
KFCV consists in dividing the data  into  $k$ subsets. Each subset plays the role of validation set while the remaining  $k-1$ subsets are used together as the training set. LOO-CV method is a particular case of KFCV with $k=n$.

The sub-models $\hat{s}_{n,-i}$ introduced in  paragraph \ref{sec:form} are used  to compute LOO estimator of the master surrogate model $ \hat{s}_n$. In fact, the  LOO errors  are $\varepsilon_i =  \hat{s}_{n,-i}(\mathbf{x_i}) - y_i$.  Notice that the sub-models  are used to estimate a feature of the master surrogate model. In our study, we will be interested in the distribution of the local predictor for all  $ \mathbf{x}  \in \mathbb{X}$ ($\mathbf{x}$  is not necessarily a design point) and  we will also use the sub-models to estimate this feature. Indeed, this distribution will be estimated by using LOO-CV predictions leading to the definition of the Universal Prediction (UP)  distribution.
\section{Universal  Prediction distribution} \label{sec:up}
\subsection{Overview}
 As discussed in the previous section, cross-validation is used as a method for estimating the prediction error of a given model. In our case, we introduce a novel use of  cross-validation in order to estimate the local uncertainty of a surrogate model prediction. In fact,
 we assume, in Equation \eqref{eqbin},  that CV errors are an approximation of the errors of the master model. The idea is to consider CV prediction as realizations of $ \hat{s}$.

Hence, for a given surrogate model $ \hat{s}$  and for any $\mathbf{x}\in \mathbb{X}$, % the $n$ observations
$ \hat{s}_{n,-1}(\mathbf{x}),\dots,$ $ \hat{s}_{n,-n}(\mathbf{x})$  define an empirical distribution of  $ \hat{s}(\mathbf{x})$ at $\mathbf{x}$. 
In the case of an interpolating surrogate model and a deterministic simulation code $s$,  it is natural to enforce a zero variance at design points. Consequently, when predicting on a design point $\mathbf{x_i}$ we neglect the prediction $ \hat{s}_{n,-i}$. 
This can be achieved by introducing weights on the  empirical distribution. 
These weights  avoid the  pessimistic sub-model predictions that might occur in a region while the global surrogate model fits  the data well in that region. Let 
$ \hat{F}_{n,\mathbf{x}}^{(0)}$ be the weighted empirical distribution based on the $n$ different predictions of the  LOO-CV sub-models $\{   \hat{s}_{n,-i}(\mathbf{x})\}_{1\leq i \leq n}$  and weighted by $w_{i,n}(\mathbf{x})$ defined in Equation \eqref{eqbin}:

\begin{equation} 
  w_{i,n}^{0}(\mathbf{x}) =  \left\{
  \begin{aligned}
    \frac{1}{n-1} & & \text{if }  \mathbf{x_i} \neq \arg\min\{d(\mathbf{x},\mathbf{x_i}), i=1,\dots,n\}\\
   0 & & \text{otherwise}  
  \end{aligned}
  \right.  \label{eqbin}
\end{equation}

Such binary weights lead  to unsmooth  design criteria. In order to avoid this drawback, we introduce smoothed weights. 
Direct smoothing based on convolution product would lead to computations  on  Voronoi cells.  We prefer to use  the simpler smoothed  weights defined in Equation \eqref{eq:w2}.

\begin{equation} 
 w_{i,n}(\mathbf{x}) = \frac{ 1- e^{- \frac{d((\mathbf{x},\mathbf{x_i}))^2}{\rho^2}}}{\sum\limits_{j=1}^{n} \Big(1- e^{- \frac{d(\mathbf{x},\mathbf{x_j})^2}{\rho^2}}\Big)} \label{eq:w2}
\end{equation}

Notice that $w_{i,n}(\mathbf{x})$ increases with the distance between the $i^{th}$ design point $\mathbf{x_i}$ and $\mathbf{x}$. In fact, the  least weighted predictions is  $ \hat{s}_{n,-p_{nn}(\mathbf{x})}$ where  $p_{nn}(\mathbf{x})$ is the index of the nearest design point  to $\mathbf{x}$. 
 In general, the prediction $ \hat{s}_{ n,-i}$ is locally less reliable in a neighborhood of $\mathbf{x_i}$. The proposed weights determine  the local relative confidence level of a given sub-model predictions. The term ``relative'' means that the confidence level of one sub-model prediction is relative to the remaining sub-models predictions due to the normalization factor  in Equation \eqref{eq:w2}. 
The smoothing parameter $\rho$ tunes the amount of  uncertainty  of  $  \hat{s}_{n,-i}$  in a neighborhood of $\mathbf{x_i}$. Several options are possible to choose $\rho$. We suggest setting $\rho = \bar{d}(\mathbf{X_n})$. Indeed, this is a well suited choice for practical cases.

\begin{definition}
The Universal Prediction distribution (UP distribution) is the  weighted empirical distribution: 
 \begin{equation}\label{eq:mudf}
\mu_{(n,\mathbf{x})}(dy) = \sum\limits_{i=1}^{n} w_{i,n}(\mathbf{x}) \delta_{ \hat{s}_{n,-i}(\mathbf{x})} (dy).  
\end{equation}
\end{definition}
This probability measure is nothing more than the empirical distribution of all the predictions provided  by cross-validation sub-models weighted by local smoothed masses.

\begin{definition} For  $ \mathbf{x}  \in \mathbb{X}$ we call $\hat{\sigma}^2_{n}(\mathbf{x})$  (Equation \eqref{eq:var}) the local UP variance and   $\hat{m}_{n}(\mathbf{x})$  (Equation  \eqref{eq:mean})  the UP expected value.  
  \begin{equation} \label{eq:mean}
 \hat{m}_n(\mathbf{x})= \int y  \mu_{(n,\mathbf{x})}(dy)   = \sum\limits_{i=1}^{n} w_{i,n}(\mathbf{x})  \hat{s}_{n,-i}(\mathbf{x})
 \end{equation}  
 
 \begin{equation}\label{eq:var}
\hat{\sigma}^2_{n}(\mathbf{x}) = \int (y - \hat{m}_{n}(\mathbf{x}) )^2 \,\mu_{(n,\mathbf{x})}(dy) 
  = \sum\limits_{i=1}^{n} w_{i,n}(\mathbf{x})(\hat{s}_{n,-i}(\mathbf{x})-  \hat{m}_{n}(\mathbf{x}) )^2
 \end{equation} 
 
 \end{definition}

\subsection{Illustrative example} 
 Let us consider the Viana function defined over  $[ - 3, 3 ]$  %the Equation \eqref{eq:viana}.
\begin{equation}\label{eq:viana}  
f(\mathbf{x}) = \frac{10 \cos(2x) + 15 - 5x +x^2}{50}
\end{equation}

\begin{figure}[ht] 
  \centering
  \includegraphics[width=0.7\textwidth, height = 5.5cm,natwidth=8.89in,natheight=8.40in]{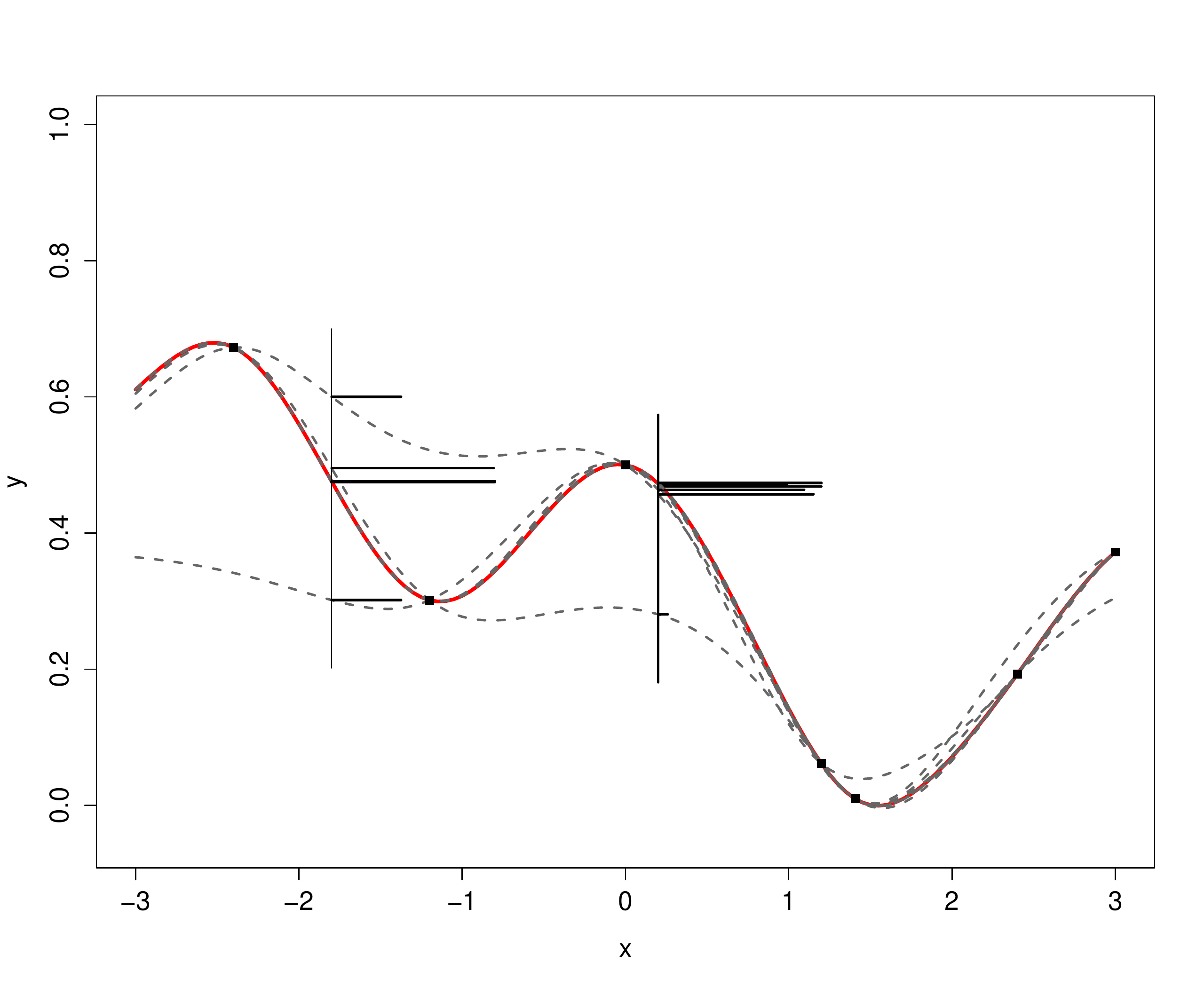}
  \caption{Illustration of the UP distribution. Dashed lines: CV sub-models predictions, solid red line: master model prediction, horizontal bars: local  UP distribution at \(x_a= - 1.8\) and \(x_b=0.2\), black squares: design points.}
 \label{fig:viana7dp}
\end{figure}

 Let $\mathbf{Z_n} =(\mathbf{X_n},\mathbf{Y_n})$ be the design of experiments such that  $\mathbf{X_n} = (x_1=-2.4,\mathbf{x}_2,=-1.2,x_3=0,x_4=1.2,x_5=1.4,x_6=2.4,x_7=3)$ and $\mathbf{Y_n}=(y_1,\dots,y_7)$ their image by $f$. 
  We used  a  Gaussian process regression   \cite{matheron1963principles,krig1,krig2} with constant trend function and  Mat\'ern 5/2 covariance function $\hat{s}$. We display in Figure \ref{fig:viana7dp}  the design points, the cross-validation sub-models predictions $\hat{s}_{n,-i}$, $i =1,\dots7$  and  the master model prediction  $\hat{s}_{n}$.

Notice that in  the  interval $[ 1, 3]$ (where we have 4 design points) the discrepancy between the master model and the CV sub-models predictions is smaller than in the remaining space. 
Moreover, we displayed horizontally the \textit{UP distribution} at $x_a= - 1.8$ and $x_b=0.2$ to illustrate the weighting effect. One can  notice that: 
\begin{itemize}
\item At $x_a$ the least weighted predictions are  $\hat{s}_{n,-1}(x_a)$ and $\hat{s}_{n,-2}(x_a)$. These predictions do not use the two closest design points to $x_a$ : ($x_1$, respectively $x_2$).
\item At $x_b$, $\hat{s}_{n,-3}(x_b)$ is the least weighted prediction.
\end{itemize}
  
\begin{figure}[ht] 
  \centering
  \includegraphics[width=0.7\textwidth,height = 5.5cm,natwidth=8.99in,natheight=8.03in]{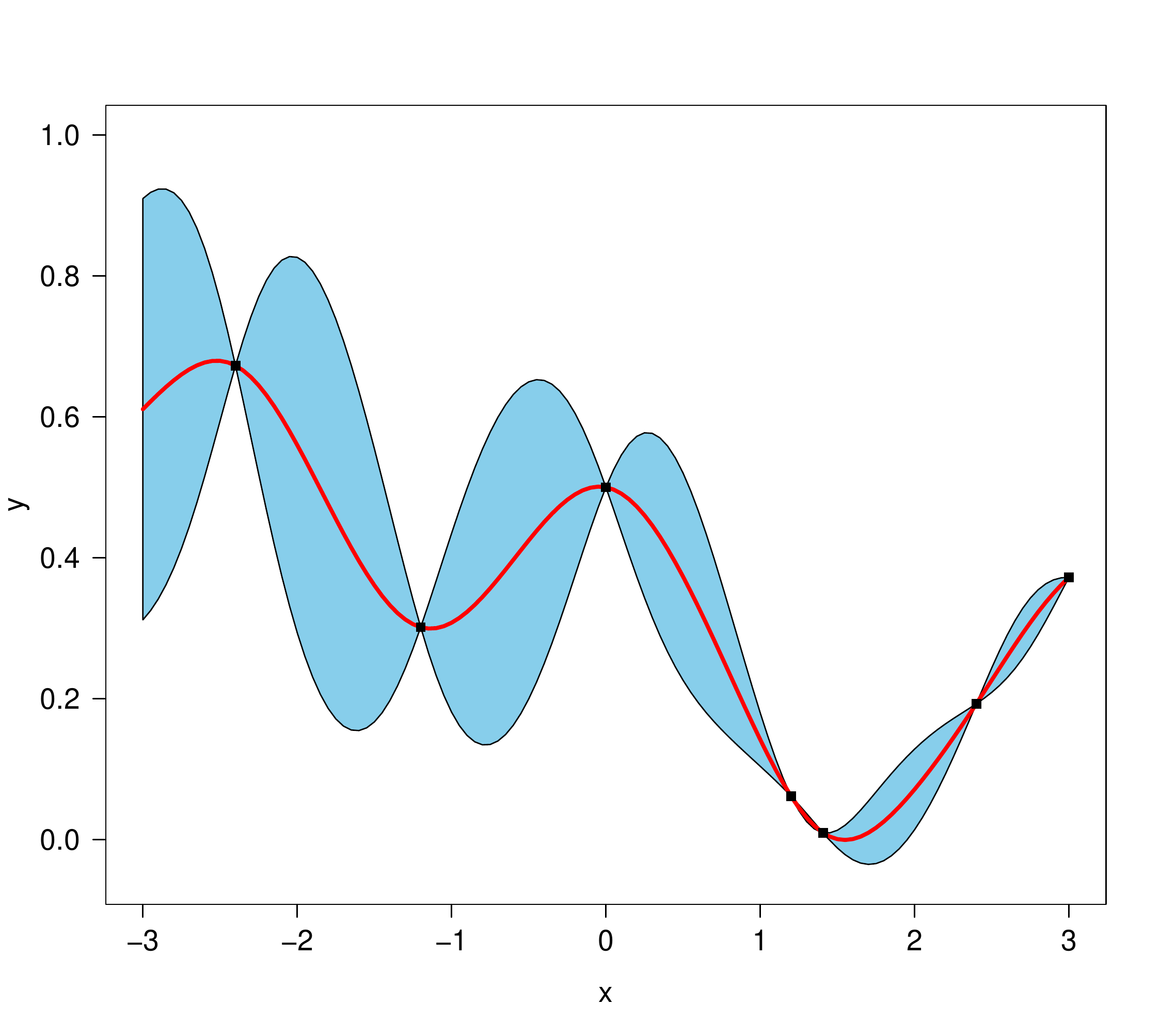}
  \caption{Uncertainty quantification based on the UP distribution. Red solid line: master model prediction $\hat{s}_{n}(\mathbf{x})$, blue area: region delimited by $\hat{s}_{n}(\mathbf{x}) \pm  3 \hat{\sigma}_{n}(\mathbf{x}) $.}
 \label{fig:Vianavar}
\end{figure}

Furthermore, we display in Figure \ref{fig:Vianavar} the master  model prediction   and region delimited  by $\hat{s}_{n}(\mathbf{x}) + 3 \hat{\sigma}_{n}(\mathbf{x}) $ and $\hat{s}_{n}(\mathbf{x}) - 3 \hat{\sigma}_{n}(\mathbf{x}) $.

One can  notice  that the standard deviation is null at design points. In addition, its local maxima  in the interval  $[ 1,3 ]$ (where we have more design points density) are smaller than its maxima in the remaining space region.

\section{Sequential Refinement} \label{sec:Ref}
In this section, we use the \textit{UP distribution}  to define an adaptive refinement technique called the Universal Prediction-based Surrogate Modeling Adaptive Refinement Technique UP-SMART.
\subsection{Introduction}
The main goal of sequential design is to minimize the number of calls of a computationally expensive function. Gaussian surrogate models \cite{krig2} are widely used in adaptive design strategies. Indeed, Gaussian modeling gives a Bayesian framework for sequential design. In some cases, other surrogate models might be more accurate although they do not provide    a theoretical framework for uncertainty assessment.  We propose here a new universal strategy for adaptive sequential design of experiments. The technique is  based on the UP distribution. So, it can be applied to any type of surrogate model.

In the literature, many strategies have been proposed to design the experiments (for an overview, the interested reader is referred to \cite{giunta2003overview,wang2007review,shao2008clustering}). Some strategies, such as Latin Hypercube Sampling (LHS)  \cite{mckay1979comparison}, maximum entropy design  \cite{shewry1987maximum}, and maximin distance designs  \cite{johnson1990minimax} are  called  one-shot sampling methods. These methods  depend neither on the output values nor on the surrogate model. However, one would naturally expect to design more points in the regions with high nonlinear behavior. %  than in space region with a nearly linear behavior. 
This intuition leads to  adaptive  strategies. A DOE approach is said to be adaptive when information from the experiments (inputs and responses) as well as information from surrogate models are used to select the location of the next point. 

By adopting this definition, adaptive DOE  methods include for instance surrogate model-based optimization algorithms, probability of failure estimation techniques and sequential refinement techniques. Sequential refinement techniques aim at creating a  more accurate surrogate model. For example,  Lin et al. \cite{lin2004sequential} use  Multivariate Adaptive Regression Splines (MARS) and kriging models with Sequential Exploratory Experimental Design (SEED) method. It consists in building a surrogate model to predict errors based on the errors on a test set.  Goel et al. \cite{goelensemble}  use an ensemble of surrogate models to identify regions of high uncertainty by computing the  empirical  standard deviation  of the predictions of the ensemble members.  
Our method is based on the predictions of the CV sub-models. In the literature, several  cross-validation-based techniques  have been discussed. Li et al. \cite{li2006maximum}  propose to  add the design point that maximizes the  Accumulative Error (AE). The AE on  $ \mathbf{x}  \in \mathbb{X}$ is computed as the sum of the LOO-CV errors %multiplied by a degree of influence factor for each prediction.
on the design points weighted by influence factors.
This method  could lead to clustered samples. To avoid this effect, the authors \cite{li} propose to add a threshold constraint in the maximization problem. Busby et al.  \cite{busby2007hierarchical} propose a method based on a grid and CV. It affects the CV prediction errors at a design point to its containing cell in the grid.
Then, an entropy approach is performed to add a new design point. More recently, Xu et al. \cite{xu2014robust}  suggest the use of a method based on Voronoi cells and CV. 
Kleijnen et al.\cite{kleijnen2004application}  propose a method based on the Jackknife's pseudo values predictions variance. Jin et al. \cite{jin2002sequential}  present a strategy  that maximizes the product between the deviation  of CV sub-models predictions with respect to the master model prediction and  the distance to the design points.
 Aute et al. \cite{aut} introduce the Space-Filling Cross-Validation Trade-off (SFCVT)  approach. It consists in building a new surrogate model over  LOO-CV errors and then add a point that maximizes the new surrogate model prediction under some space-filling constraints. %threshold distance to the initial design constraint called here space filling constraint. 
 In general cross-validation-based approaches tend to allocate points close to each other resulting in clustering  \cite{aut}. This is not desirable for deterministic simulations.

\subsection{UP-SMART} 
The idea behind UP-SMART is to sample points where the UP distribution variance (Equation \eqref{eq:var}) is maximal. Most of the  CV-based sampling criteria use   CV errors. Here, we use the local predictions of the CV sub-models. Moreover,   
notice that the UP variance is null on design points for interpolating surrogate models. Hence, UP-SMART  does not naturally promote clustering.
  
However, $ \hat{\sigma}^2_{n}(\mathbf{x})$ can vanish even if $\mathbf{x}$ is not a design points. To overcome this drawback, we add a distance penalization. This leads to  the UP-SMART sampling criterion   $\gamma_{n}$ (Equation  \eqref{eq:critRP}).
 \begin{equation}\label{eq:critRP}
  \gamma_{n}(\mathbf{x})  =  \hat{\sigma}^2_{n}(\mathbf{x}) + \delta  \underline{d}_{\mathbf{X_n}} (\mathbf{x})
 \end{equation} 
  where  $\delta>0$ is called  exploration parameter. One can set $\delta$  as a small percentage of the global variation of the output. 
 UP-SMART is the adaptive refinement algorithm consisting in  adding at step $n$ a point
 $x_{n+1} \in \arg \max\limits_{   \mathbf{x}   \in \mathbb{X}} (\gamma_{n}(\mathbf{x}))$.
 
\subsection{Performances on a set of test functions}
In this subsection, we present the performance of the UP-SMART. We  present first the used  surrogate-models.
\subsubsection{Used surrogate models}
\paragraph{Kriging}
Kriging ~\cite{matheron1963principles} or Gaussian process regression is an interpolation method.  

Universal Kriging fits the data using a deterministic trend  and governed by prior covariances.  
Let  $ k(\mathbf{x},\mathbf{x'})$,  be  a covariance function on  $\mathbb{X} \times \mathbb{X}$,  and let $(h_i)_{1\leq i\leq p}$  be the basis functions of the trend. Let us denote $\mathbf{h}(\mathbf{x})$ the vector $(h_1(\mathbf{x}),..,h_p(\mathbf{x}))^\top$  and  let  $H$  be the matrix  with entries $h_{ij} = h_j(\mathbf{x_i}),    1\leq i, j \leq  n$. 
 Furthermore, let $\mathbf{k_n}(\mathbf{x})$ be the vector  $(k(\mathbf{x},\mathbf{x_1}),..,k(\mathbf{x},\mathbf{x_n}))^\top $ and $K_n$ the matrix with entries $ 
k_{i,j} = k(\mathbf{x_i},\mathbf{x_j})$, for $   1\leq i, j \leq  n$. 

Then, the conditional mean  of the Gaussian process with covariance $k(\mathbf{x},\mathbf{x'})$  and  its variance are given in Equations (\eqref{eq:krigingMean},\eqref{eq:KrigingVariance})
\begin{align}
 m_{G_n}(\mathbf{x}) &= \mathbf{h}(\mathbf{x})^\top \hat{\beta} +  \mathbf{ k_n}(\mathbf{x})^\top K_n^{-1}(Y-H^\top \hat{\beta}) \label{eq:krigingMean}\\
\sigma^2_{GP_n}(\mathbf{x}) &= k(\mathbf{x},\mathbf{x})- \mathbf{ k_n}(\mathbf{x})^\top K_n^{-1}  \mathbf{ k_n}(\mathbf{x})^\top +\mathbf{V}(\mathbf{x})^\top(H^\top K_n^{-1}H)^{-1} \mathbf{V}(\mathbf{x}) \label{eq:KrigingVariance}\\
 \hat{\beta} &= (H^\top K_n^{-1}H)^{-1} H^\top K_n^{-1}Y  \text{ and } \mathbf{V}(\mathbf{x}) = \mathbf{h}(\mathbf{x})^\top+ \mathbf{ k_n}(\mathbf{x})^\top K_n^{-1}H
\end{align}

Note that the conditional mean is the prediction of the Gaussian process regression.
Further, we used two kriging instances with different sampling schemes  in our test bench. Both  use constant trend function and a  Mat\'ern 5/2 covariance function. The first design  is obtained by maximizing  the \textit{UP distribution}  variance (Equation \eqref{eq:var}). And  the second one  is obtained by  maximizing the kriging variance $\sigma^2_{GP_n}(\mathbf{x})$.%as refinement criterion. 

\paragraph{Genetic aggregation}
The genetic aggregation response surface is a method  that aims at selecting the best response surface for a given  design of experiments. It solves several surrogate models, performs  aggregation and selects the best response surface according to the cross-validation errors.   %[citeRSGA]. 

The use of such response surface, in this test bench, aims at checking the universality of the \textit{UP distribution}:  the fact that it can be applied for all types of surrogate models. 

\subsubsection{ Test bench}
In order to test the performances of the method we launched different  refinement processes for the following  set of test functions:
\begin{itemize}
\item Branin: $f_{b}(x_1,x_2)=(x_2- (\frac{5.1}{4\pi^2} )x_{1}^{2} + (\frac{5}{\pi})x_1 - 6 )^2 + 10(1-(\frac{1}{8\pi}))\cos(x_1)+10$. 
\item Six-hump camel:  $f_c(x_1,x_2) = \left(4-2.1x_1^2+\frac{x_1^4}{3}\right)x_1^2 + x_1x_2+x_2^2(4x_2^2-4) $.
\item Hartmann6: $f_h(\mathbf{X}=(x_1,\dots,x_6)) = -\sum\limits_{i=1}^4 \alpha_i \exp\Big(- \sum\limits_{j=0}^6 A_{ij} (x_j - P_{ij})^2 \Big) $. $A$,$P$ and $\alpha$ can be found in \cite{dixon1978towards}.
\item Viana: (Equation \eqref{eq:viana})
\end{itemize}

For each function we generated by optimal Latin hyper sampling design  the number of initial design points $n_0$, the number of refinement points $N_{max}$. We also generated a set of $N_t$ test points and their response $Z^{(t)} = (X^{(t)},Y^{(t)})$. The used values are available in Table \ref{tab:1}. 

\begin{table}[ht]
\begin{center}
\caption{Used test functions}
\label{tab:1} 
\begin{tabular}{p{0.2\textwidth}p{0.15\textwidth}p{0.2\textwidth}p{0.15\textwidth}p{0.1\textwidth}} 
\hline
Function & dimension {\rm $d$}&  {\rm $n_0$} &  {\rm $N_{max}$} & {\rm $N_t$}\\
\hline
Viana  & 1 & 5 & 7&500\\
%"W function & 2 & 5 &20\\
Branin & 2 & 10&10 &1600\\
%Ackley & 2 & 15 &40\\
%Sasena \cite{sasena1998optimization} & 2 & 20 &50\\
%Extended Rosenbrock  \cite{rosenbrock1960automatic} & 4 & 25 &50\\
Camel   & 2 & 20 &10 &1600\\
Hartmann6   & 6 & 60 &150 &10000\\
%\noalign{\smallskip}
\hline
\end{tabular}
\end{center}
\end{table}

We fixed $n_0$ in order to get  non-accurate surrogate models at the first step. Usually, one follows the  rule-of-thumb  $n_0=10\times d$ proposed in \cite{loeppky2009choosing}. However, for Branin and Viana functions, this rule leads to a very good initial fit. Therefore, we choose lower values.  %Therefore, for Viana and Branin functions, we use  a lower value than the  rule-of-thumb  $n_0=10\times d$ proposed in \cite{loeppky2009choosing}. Further, we launched  the following three refinement processes for the set of test functions:
\begin{itemize}
\item Kriging  variance-based refinement process (Equation  \eqref{eq:KrigingVariance}) as refinement criterion. 
\item Kriging  using the  UP-SMART: UP-variance as refinement criterion (Equation  \eqref{eq:critRP}).
\item Genetic aggregation  using the  UP-SMART: UP-variance as refinement criterion (Equation  \eqref{eq:critRP}). %UP-variance as refinement criterion (equation  \eqref{eq:critRP} ).
\end{itemize}

\subsubsection{Results}
\begin{figure}[!ht]
 \centering \subfloat[][Viana Function]{ \includegraphics[width=0.49\textwidth,height = 5cm,natwidth=10in,natheight=5.86in] {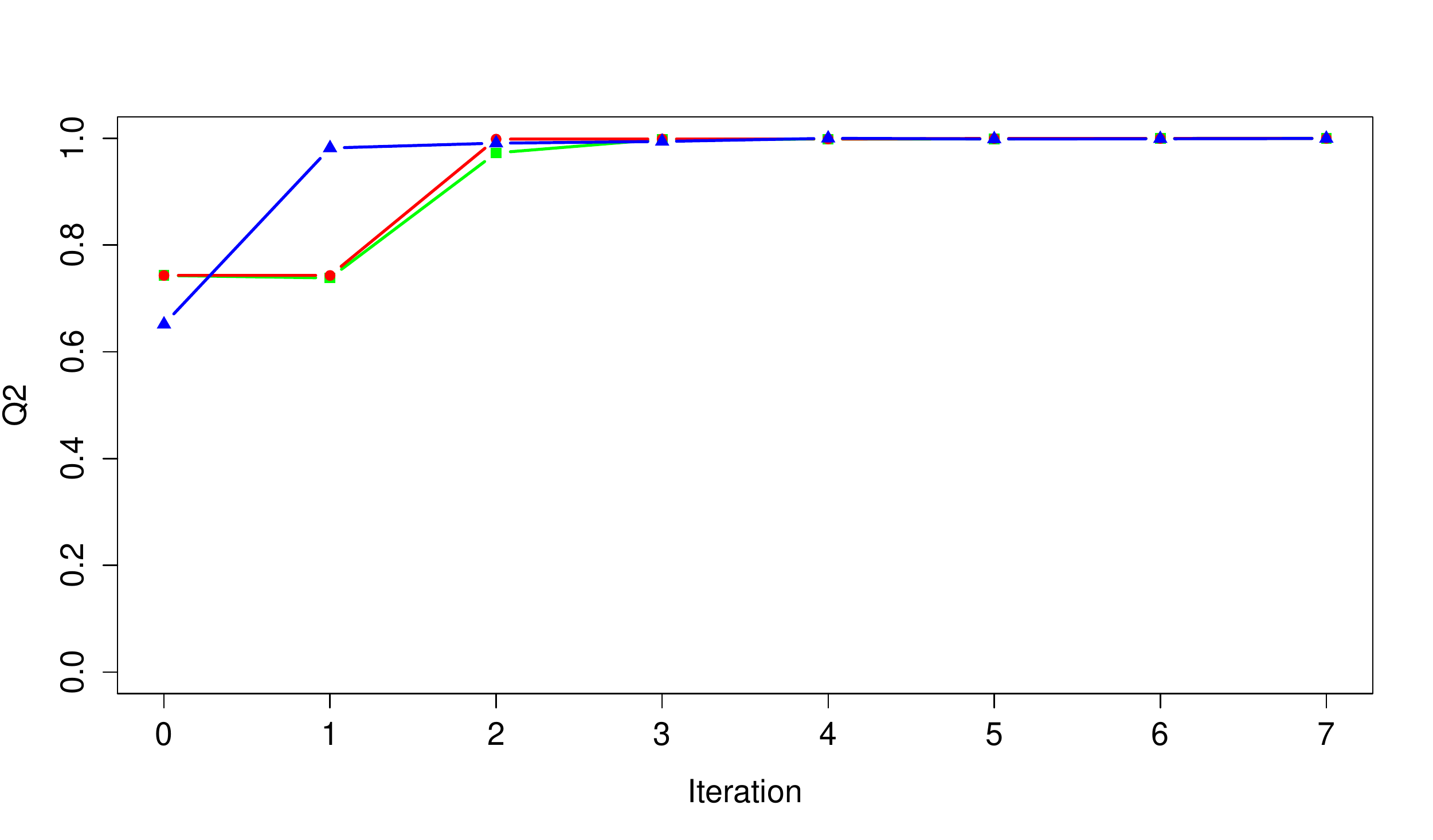}\label{fig:RP1}} \par\medskip
   \subfloat[][Branin function]{ \includegraphics[width=0.49\textwidth,height = 5cm,natwidth=10in,natheight=5.86in]{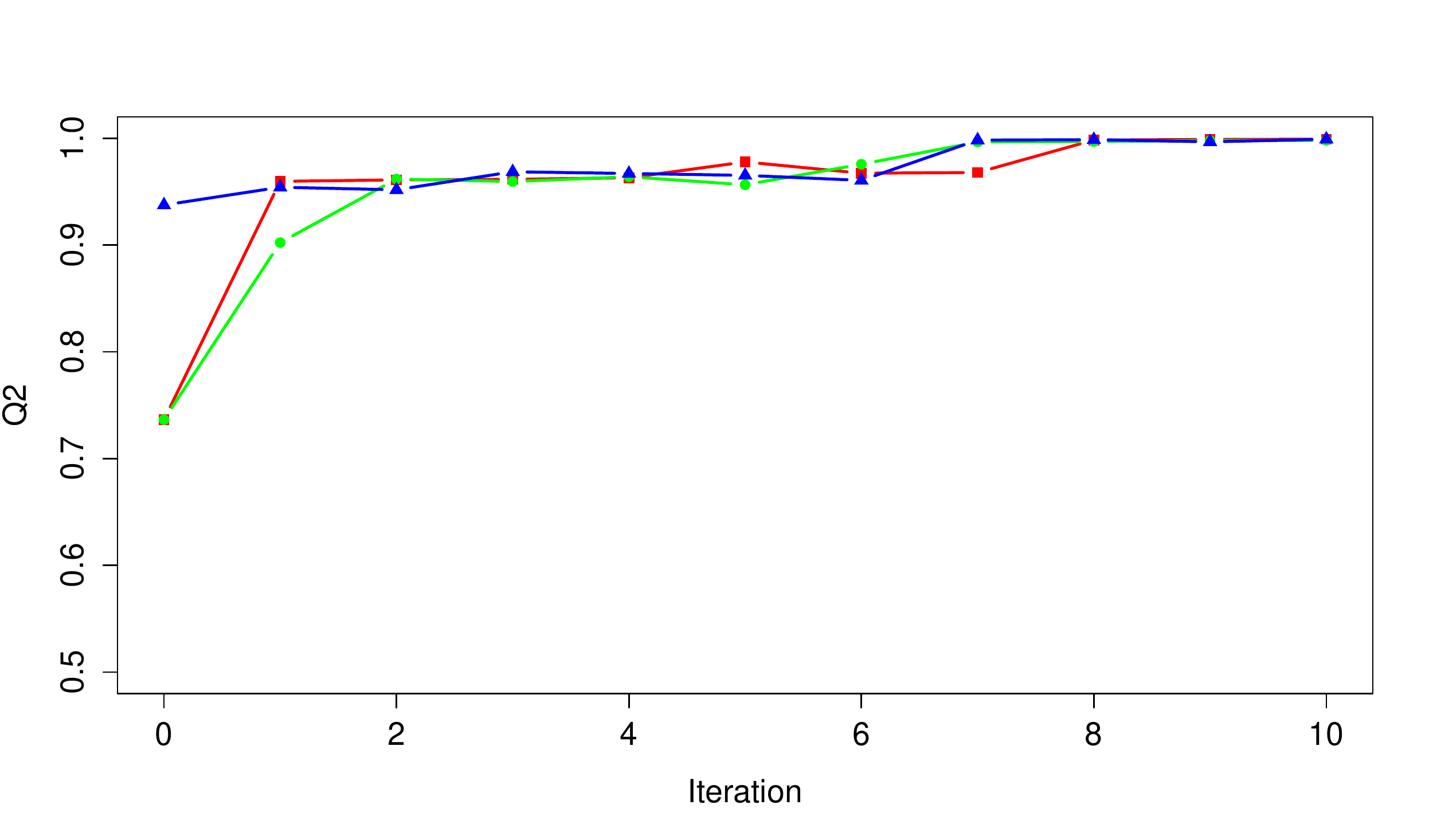}\label{fig:RP2}} 
   \subfloat[][Camel function]{\includegraphics[width=0.49\textwidth,height = 5cm,natwidth=10in,natheight=5.86in]{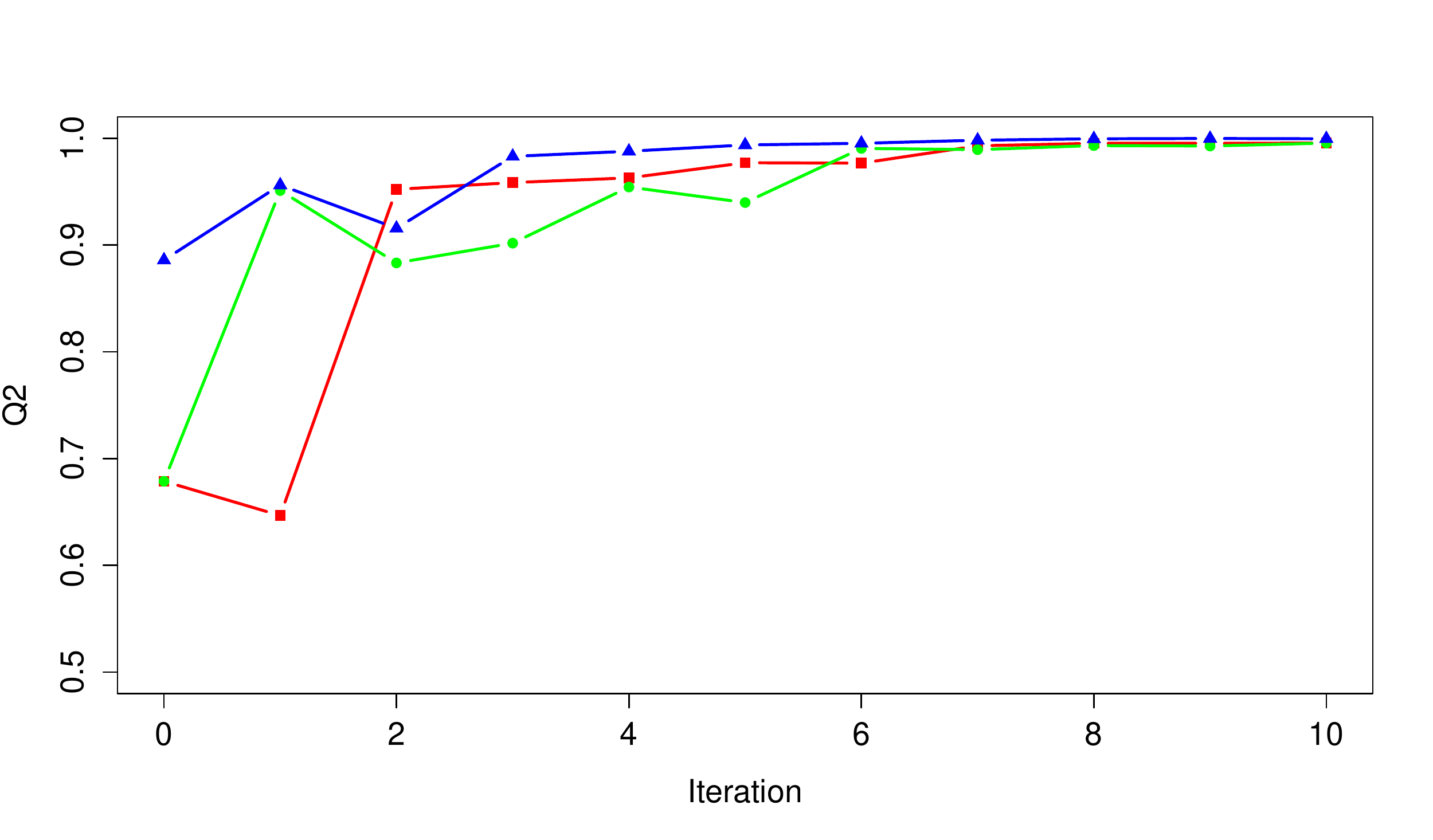}\label{fig:RP3}}%\label{fig:RP3}
  \caption{Performance of three refinement strategies on three test functions measured by the {\rm $Q^2$}  criterion on a test set. x axis: number of added refinement points. y axis: {\rm $Q^2$}. UP-SMART with kriging in green,  UP-SMART with genetic aggregation in blue and  kriging variance-based technique in red.}  \label{fig:resRP}
\end{figure}

 For each function, we compute at each iteration the Q squared ($Q^2$) of the predictions of the test set $Z^{(t)} $ where $ Q^2(\hat{s}, Z^{(t)})  = 1 -  \frac{\sum\limits_{i=1}^{N_t}(y_i^{(t)} - \hat{s}(\mathbf{x_i}^{(t)}))^2}{\sum\limits_{i=1}^{N_t} (y_i^{(t)} - \bar{y})^2} $
 and $ \bar{y} = \frac{1}{N_t}\sum\limits_{i=1}^{N_t} y_i^{(t)}  $.  
 We display in Figure \ref{fig:resRP} the performances  of  the three different techniques described above  for Viana (Figure \ref{fig:RP1}), Branin (Figure \ref{fig:RP2})  and Camel (Figure  \ref{fig:RP3}) functions measured by $Q^2$ criterion. 

\begin{figure}[!ht]
  \centering
    \includegraphics[width=0.6\textwidth,height = 5cm,natwidth=12.49in,natheight=8.39in]{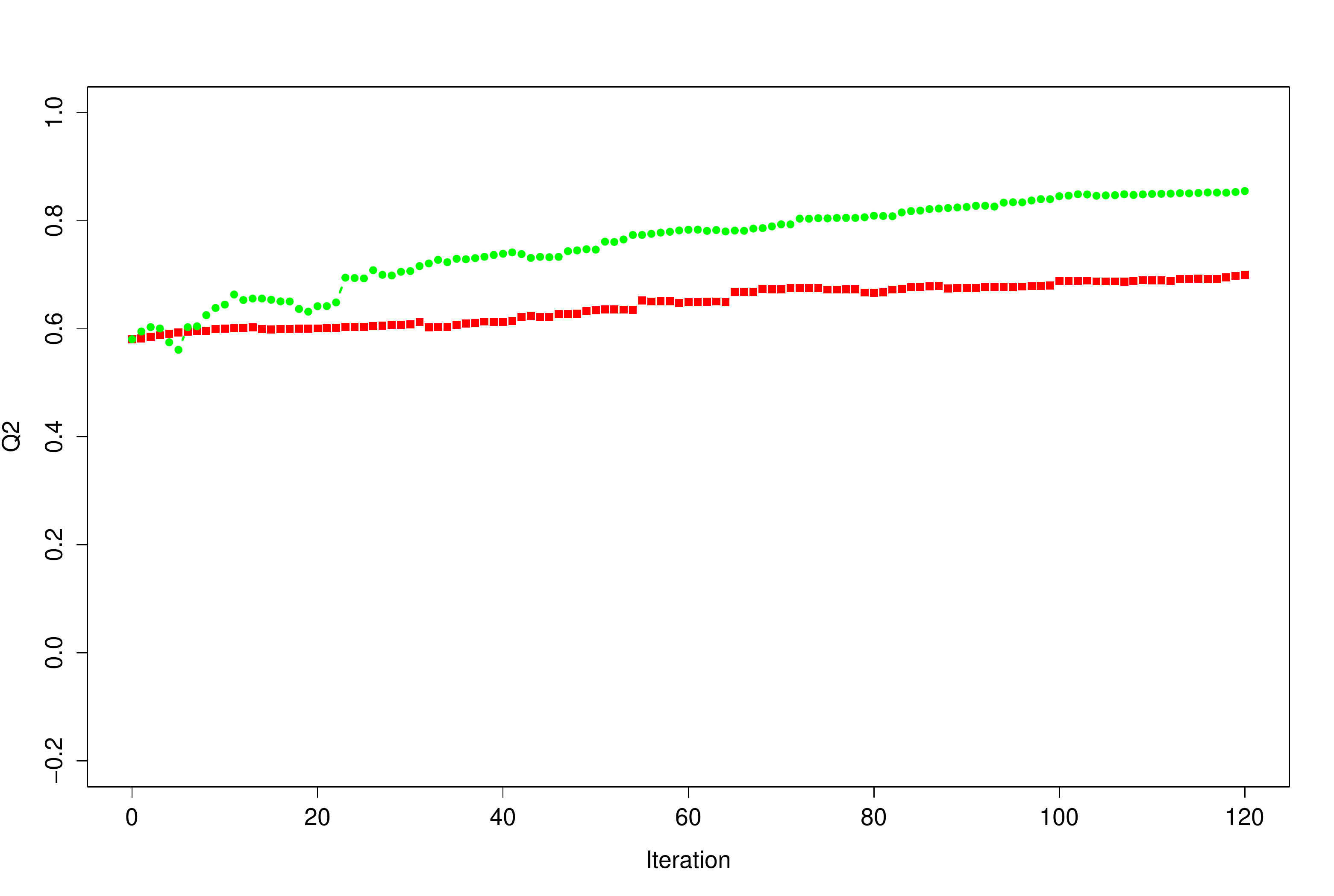}
    \caption{Performance of two refinement strategies on Hartmann function measured by the {\rm $Q^2$}  criterion on a test set. x axis: number of added refinement points. y axis: {\rm $Q^2$}. UP-SMART with kriging in green and  kriging variance-based technique in red.}
    \label{fig:hartmannRP}
\end{figure}

For these tests,  the  three techniques have comparable performances. The $Q^2$ converges for all of them.  It appears that the UP variance criterion refinement process gives  at least as good a result as the kriging variance criterion. This may be due to the high kriging uncertainty on the boundaries. In fact, minimizing kriging variance sampling algorithm generates, in general, more points on the boundaries for a high dimensional problem. For instance, let us focus on the  Hartmann function (dimension 6).  We present, in Figure \ref{fig:hartmannRP},  the  results after 150 iterations of the algorithms. It is clear that the UP-SMART gives a better result for this function. 

 The results show that: 
 \begin{itemize}
 \item %First, let us focus on the two kriging based refinement processes.  The one that uses 
 UP-SMART gives a better global response surface accuracy than the maximization of the variance. This shows the usefulness of the method. 
 \item UP-SMART is a universal  method. Here, it has been applied with success to an aggregation of response surfaces. Such usage highlights the universality of the strategy.
 \end{itemize}
 
\section{Empirical Efficient Global optimization} \label{sec:Opt}
In this section, we introduce \textit{UP distribution}-based Efficient Global Optimization (UP-EGO) algorithm. This algorithm is an adaptation of the well known EGO algorithm.  
\subsection{Overview}
Surrogate model-based optimization refers to the idea of speeding optimization processes
using surrogate models. In this section, we present an adaptation of the well-known efficient global optimization (EGO) algorithm  \cite{jonesEGO}. Our method is based on the weighted empirical distribution \textit{UP distribution}. We show that asymptotically, the points generated by the algorithm are dense around the optimum. For the EGO al The proof has been done by Vazquez et al. \cite{vazquezconvergence} for the EGO algorithm.

The basic unconstrained  surrogate model-based optimization  scheme can be summarized as follows  \cite{queipo2005surrogate}
\begin{itemize}
\item Construct a surrogate model from a set of known data points.
\item Define a sampling criterion that reflects a possible improvement.
\item Optimize the criterion over the design space.
\item Evaluate the true function at the criterion optimum/optima.
\item Update the surrogate model using new data points.
\item Iterate until convergence
\end{itemize}
Several sampling criteria have been proposed to  perform optimization. The Expected Improvement (EI) is one of the most popular criteria for surrogate model-based optimization. Sasena et al. \cite{sasena2000metamodeling}  discussed some  sampling criteria such as the threshold-bounded extreme, the regional extreme, the generalized expected improvement and the minimum surprises criterion.  Almost all of the criteria are computed in practice within  the frame of Gaussian processes. Consequently,  among all possible response surfaces, Gaussian surrogate models are widely used in surrogate model-based  optimization. Recently,  Viana et al. \cite{msego}  performe multiple surrogate assisted optimization by importing Gaussian uncertainty estimate.  
\subsection{ UP-EGO Algorithm}
Here, we use  the \textit{UP distribution} to compute an empirical expected improvement. Then,  we present an optimization  algorithm similar to the original EGO algorithm that can be applied with any type of surrogate models.
Without loss of generality,  we consider the minimization  problem: 
\begin{equation*}
\begin{aligned}
  \underset{\mathbf{x}\in \mathbb{X}}{\text{minimize}}& &
    s(\mathbf{x})  
    \end{aligned}
\end{equation*}
Let $(y(\mathbf{x}))_{\mathbf{x}\in \mathbb{X}}$ be a Gaussian process model. 
  $ m_{G_n}$ and $\sigma^2_{GP_n}$  denote  respectively the mean and the variance of the conditional process $y(\mathbf{x})\mid \mathbf{Z_n}$.  Further, let $ y_n^\star$ be the  minimum value at  step  $n$ when using observations $\mathbf{Z_n}=(z_1,\dots,z_n)$ where $z_i=(\mathbf{x_i},y_i)$. 
($y_n^\star = \min\limits_{i=1..n}{y_i}$).
 The EGO algorithm \cite{jonesEGO}  uses  the  expected improvement  $EI_n$ (Equation \eqref{eq:EGO}) as sampling criterion: 
  \begin{equation}
 \label{eq:EGO}
  EI_n(\mathbf{x})  =  \mathbb{E} [ \max(y_n^\star- y(\mathbf{x}),0 ) \mid \mathbf{Z_n}]
 \end{equation} 
  The EGO algorithm adds the point that maximizes  $EI_n$  . Using some Gaussian computations,   Equation \eqref{eq:EGO} is equivalent to  Equation \eqref{eq:EI}.
 \begin{equation}
 \label{eq:EI}
EI_n(\mathbf{x})  = \left\{
  \begin{aligned}
 (y_n^\star -   m_{G_n}(\mathbf{x}))\Phi \left(\frac{ y_n^\star-  m_{G_n}(\mathbf{x})}{\sigma_{GP_n}(\mathbf{x})}\right) +\sigma_{GP_n}(\mathbf{x}) \phi\left(\frac{ y_n^\star-  m_{G_n}(\mathbf{x})}{\sigma_{GP_n}(\mathbf{x})}\right)  & & \text{if } \sigma_{GP_n}(\mathbf{x}) \neq 0\\
0 & &\text{otherwise}\\
  \end{aligned}
  \right.  
 \end{equation}
 We introduce a similar criterion  based on the \textit{UP distribution}. With the notations of Sections \ref{sec:notations} and \ref{sec:up}, $EEI_n$ (Equation \eqref{eq:EEI})  is called the  empirical expected improvement. 
 \begin{equation}
 \label{eq:EEI}
 \begin{aligned}
  EEI_n(\mathbf{x})  &=  \int \max(y_n^\star- y,0)   \mu_{(n,\mathbf{x})}(dy)\\   
  &= \sum\limits_{i=1} w_{i,n}(\mathbf{x}) \max(y_n^\star - \hat{s}_{n,-i}(\mathbf{x}),0 )
 \end{aligned}
 \end{equation} 
We can remark that $EEI_n(\mathbf{x})$ can vanish even if $\mathbf{x}$ is not a design point.  This is  one of  the limitations of the  empirical \textit{UP distribution}. To overcome this drawback, we suggest the use of the  Universal Prediction Expected Improvement (UP-EI) $\kappa_n$ (Equation \eqref{eq:upei} )
 \begin{equation}
 \label{eq:upei}
  \kappa_n(\mathbf{x}) = EEI_n(\mathbf{x}) + \xi_n(\mathbf{x})
 \end{equation} 
  where $\xi_n(\mathbf{x})$ is a distance penalization. We use $\xi_n(\mathbf{x}) = \delta \underline{d}_{\mathbf{X_n}} (\mathbf{x})$ where  $\delta > 0 $ is  called the exploration parameter. One can set $\delta$  as a small percentage of the global variation of the output for less exploration. Greater value of $ \delta $ means more exploration. $\delta$ fixes the wished trade-off between exploration and local search. 
  
Furthermore, notice that $\kappa_n$ has the desirable property also verified by the usual EI:
 \begin{proposition} \label{prop:kappanull}
 $\forall n > 1, \forall \mathbf{Z_n}=(\mathbf{X_n}=(\mathbf{x_1},\dots,\mathbf{x_n})^\top,\mathbf{Y_n} = s(\mathbf{X_n})),$ if the used model interpolates the data then $ \kappa_n(\mathbf{x_i}) = 0$, for $i=1,\dots,n$  
 \end{proposition}

  The \textit{UP distribution}-based Efficient Global  Optimization (UP-EGO) (Algorithm \ref{algo:EEIKrigOpt} ) consists in sampling  at each iteration the point that maximize $\kappa_n$. The point is then added to the set of observations and the surrogate model is updated. 
 \begin{algorithm} 
 UP-EGO(${ \hat{s}}$)\\
$\text{ }\hspace{15pt} $\textbf{Inputs:}    $\mathbf{Z}_{n_0} = (X_{n_0},Y_{n_0})$, $n_0 \in \mathbb{N}\setminus \{ 0,1 \}$ and a deterministic function $s$\\
$\text{ }\hspace{15pt}(1)$ $m:= n_0$, $\mathbf{S}_m:= X_{n_0}  $,  $Y_m:= Y_{n_0} $\\
$\text{ }\hspace{15pt}(2)$ Compute the  surrogate model $ \hat{s}_{ \mathbf{Z}_{m}}$\\
$\text{ }\hspace{15pt}(3)$ \textit{Stop\_conditions} $:=$ False\\
$\text{ }\hspace{15pt}(4)$\textbf{While } Stop\_conditions are not satisfied
 
$\text{ }\hspace{40pt}(4.1)$ Select  $\mathbf{x_{m+1}} \in arg \max\limits_{\mathbb{X}}(\kappa_m(\mathbf{x}) )$

$\text{ }\hspace{40pt}(4.2)$ Evaluate  $y_{m+1}:= s(\mathbf{x_{m+1}}) $ 

$\text{ }\hspace{40pt}(4.3)$ $ \mathbf{S}_{m+1}:= \mathbf{S}_m\cup \{ \mathbf{x_{m+1}} \}$,  $Y_{m+1}:= Y_m \cup \{y_{m+1} \}$

$\text{ }\hspace{40pt}(4.4)$ $ \mathbf{Z}_{m+1}:= (\mathbf{S}_{m+1},Y_{m+1})$,$ m:= m+1$

$\text{ }\hspace{40pt}(4.5)$ Update the  surrogate model

$\text{ }\hspace{40pt}(4.6)$ Check  \textit{Stop\_conditions}\\
$\text{ }\hspace{25pt}$\textbf{end loop}\\
$\text{ }\hspace{10pt} $ \textbf{Outputs:}  $ \mathbf{Z}_{m}:= (\mathbf{S}_{m},Y_{m})$, surrogate model $ \hat{s}_{ \mathbf{Z}_{m}}$\\
\label{algo:EEIKrigOpt}
\end{algorithm}

\subsection{ UP-EGO convergence} \label{sec:upegoconv}
We first recall the context.  $\mathbb{X}$ is a nonempty compact subset of the Euclidean space $\mathbb{R}^p$  where $p\in \mathbb{N}^\star$. $s$ is an expensive-to-evaluate function. The weights of  the \textit{UP distribution} are computed as in  Equation \eqref{eq:w2} with  $\rho>0$ a fixed real parameter. Moreover, we consider the asymptotic behaviour of the algorithm so that, here, the number of iterations goes to infinity.  %We suppose that \textit{Stop\_conditions} are never satisfied. 

Let $ \mathbf{x^\star} \in \argmin\{s(\mathbf{x}),    \mathbf{x}   \in \mathbb{X}\}$  and   $ \hat{s}$ be a continuous interpolating  surrogate model bounded  on $\mathbb{X}$. 
 Let  $\mathbf{Z}_{n_0}=(X_{n_0}=  (\mathbf{x_1},\dots,\mathbf{x_ {n_0}})^\top,Y_{n_0}) $  be the initial data. For all $k > n_0 $, 
 $\mathbf{x_{k}}$  denotes  the point  generated by the UP-EGO algorithm at step  $k - n_0$.  Let  $\mathbf{S}_m$ denote the set  $ \{ \mathbf{x_i}, i\leq m \}$ and $S= \{ \mathbf{x_i}, i>0 \}$. Finally,  $\forall m > n_0$ we note $\kappa_{m}$ the UP-EI  of $\hat{s}_{\mathbf{Z}_{m}}$.
We are going to prove that $ \mathbf{x^\star}$ is adherent to the sequence $S$ generated by the UP-EGO($\hat{s}$) algorithm.

\begin{lemma} \label{prop:wu}
%$\forall n_0>1$, Let $\mathbf{Z}_{n_0}=(X_{n_0},Y_{n_0})$  be the initial  data. 
$\exists  \theta >0$, $\forall m \geq n_0$,  $ \forall \mathbf{x} \in \mathbb{X}$,  $\forall i \in 1,\dots,m$,
 $ \forall n > m $,  $w_{i,n}(\mathbf{x}) \leq \theta   d(\mathbf{x},\mathbf{x_i})^2 $.
\end{lemma}

\begin{definition} \label{def:interpolating}
 A surrogate model $\hat{s}$  is called an  interpolating surrogate model if for all $n \in \mathbb{N}^\star$ and for all $\mathbf{Z_n}=(\mathbf{X_n},\mathbf{Y_n}) \in \mathbb{X}^n \times \mathbb{R}^n$, $\hat{s}_{\mathbf{Z_n}}(\mathbf{x})= s(\mathbf{x})$ if $\mathbf{x} \in \mathbf{X_n}$.
\end{definition}

\begin{definition} \label{def:bounded}
 A surrogate model $\hat{s}$  is called bounded  on $\mathbb{X} $ if for all $s$ a continuous function on $\mathbb{X}$, $\exists L, U $, such that for all $n>1$ and for all  $\mathbf{Z_n}=(\mathbf{X_n},\mathbf{Y_n} = s(\mathbf{X_n})) \in \mathbb{X}^n \times \mathbb{R}^n$,   $\forall \mathbf{x}\in \mathbb{X} $,  $ L \leq \hat{s}_{\mathbf{Z_n}}(\mathbf{x})\leq U$ 
\end{definition}
 
\begin{definition} \label{def:continuous}
A surrogate model $\hat{s}$  is called  continuous if  $ \forall n_0 >1$   $\forall \mathbf{x} \in \mathbb{X}$
$\forall \varepsilon>0$, $  \exists \delta >0$, $\forall n \geq n_0$, $\forall \mathbf{Z_n}=(\mathbf{X_n},\mathbf{Y_n}) \in \mathbb{X}^n \times \mathbb{R}^n$,  $\forall \mathbf{x'}\in \mathbb{X} $,  $d(\mathbf{x},\mathbf{x'} )< \delta \implies \left|\hat{s}_{\mathbf{Z_n}}(\mathbf{x} ) - \hat{s}_{\mathbf{Z_n}}(\mathbf{x'} )  \right|<\varepsilon$
\end{definition}

\begin{theorem} \label{propconv}
Let $s$ be a  real  function defined on  $\mathbb{X}$ and let $ \mathbf{x^\star} \in \arg\min\{s(\mathbf{x}),    \mathbf{x}   \in \mathbb{X}\}$. If $ \hat{s}$ is an interpolating   continuous surrogate model  bounded  on $\mathbb{X} $, then $ \mathbf{x^\star}$ is adherent to the sequence of points $S$ generated by UP-EGO(${ \hat{s}}$).  
\end{theorem}

The proofs (Section \ref{sec:proofs}) show  that the exploration parameter is important for this theoretical result. In our implementation, we scale the input spaces to be the hypercube $ \big[-1,1] $ and we set $\delta$ to $0.005\%$ of the output variation. Hence, the exploratory effect only slightly impacts  the UP-EI criterion in practical cases.

\subsection{Numerical examples} 
Let us consider the  set of test functions (Table \ref{tab:OptFun}).

\begin{table}[ht]
\caption{Optimization test functions}
\label{tab:OptFun} % Give a unique label
\begin{center}
\begin{tabular}{p{0.2\textwidth}p{0.18\textwidth}p{0.17\textwidth}p{0.15\textwidth}}
\hline%\noalign{\smallskip}
\hline
function $f^{(i)}$ & Dimension $d^{(i)}$ & Number of initial points $n_0^{(i)} $ & Number of iterations $N_{max}^{(i)}$\\
%\noalign{\smallskip}
\hline%\noalign{\smallskip}
Branin  & 2 & 5 &40\\
Ackley  & 2 & 10 &30\\
Six-hump Camel  & 2 & 10 &30\\
%extended Rosenbrock \cite{rosenbrock1960automatic} & 4  & 40 &30\\
Hartmann6  & 6 & 20 &40\\
%\noalign{\smallskip}
\hline
\end{tabular}
\end{center}
\end{table}

We launched the optimization process for these functions with three different optimization algorithms: 
\begin{itemize}
\item EGO \cite{jonesEGO}: Implementation of the R package DiceOptim \cite{roustant2012dicekriging} using the default parameters.
\item  UP-EGO algorithm applied to a universal kriging surrogate model $\hat{s_k}$ that uses Mat\'ern 5/2  covariance function and a constant trend function. We denote this algorithm UP-EGO($\hat{s_k}$)
 \item  UP-EGO algorithm applied to the genetic aggregation $\hat{s_a}$. It is then denoted UP-EGO($\hat{s_a}$).% [citeRSGA]
\end{itemize}

\begin{figure}[!ht]
 \subfloat[][Branin ]{ \includegraphics[width=0.45\textwidth,height = 5cm,natwidth=11.75in,natheight=8.61in]{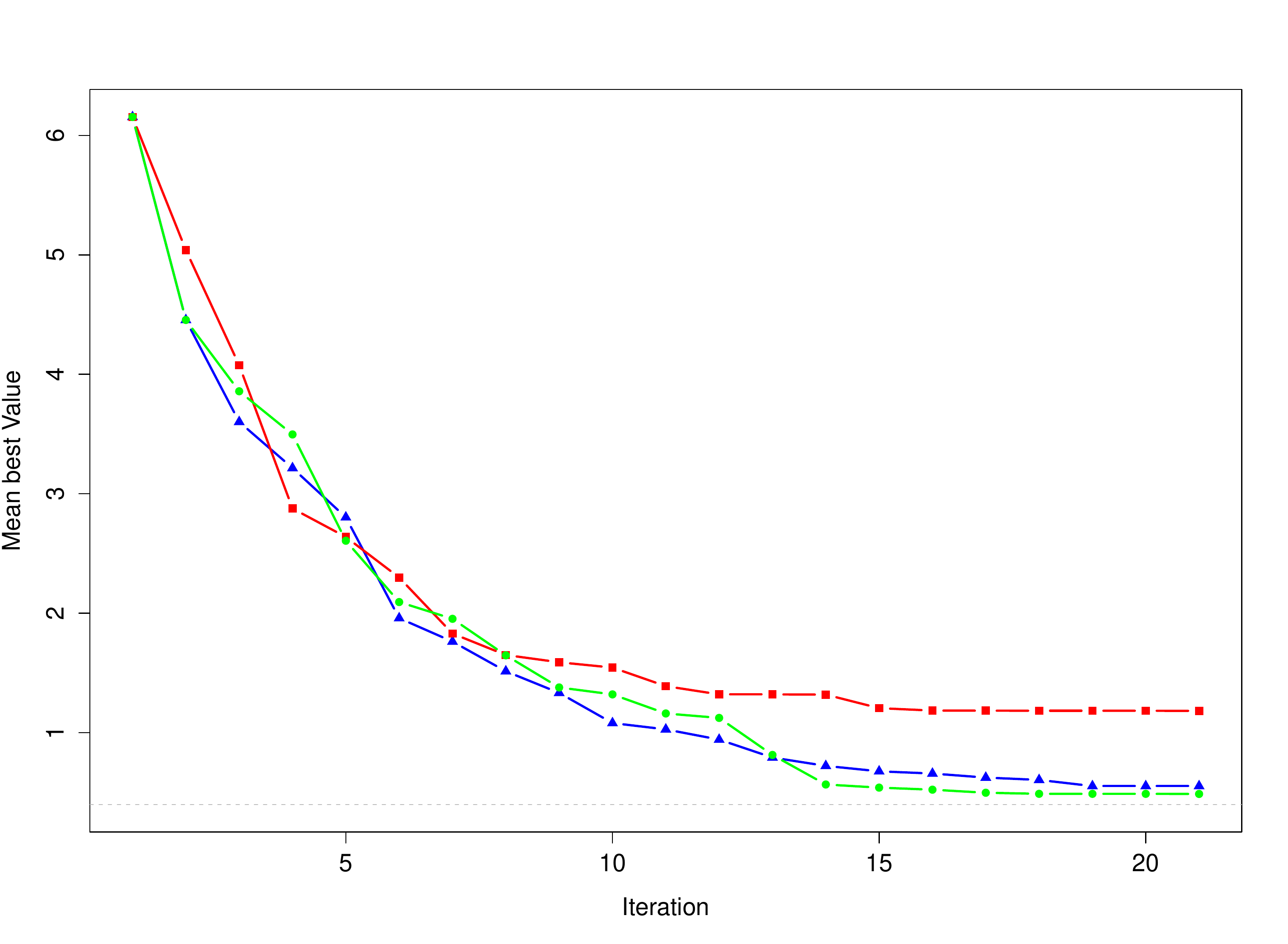}  \label{fig:opt1} } %  }
    \subfloat[][Camel]{ \includegraphics[width=0.45\textwidth,height = 5cm,natwidth=12.61in,natheight=8.61in]{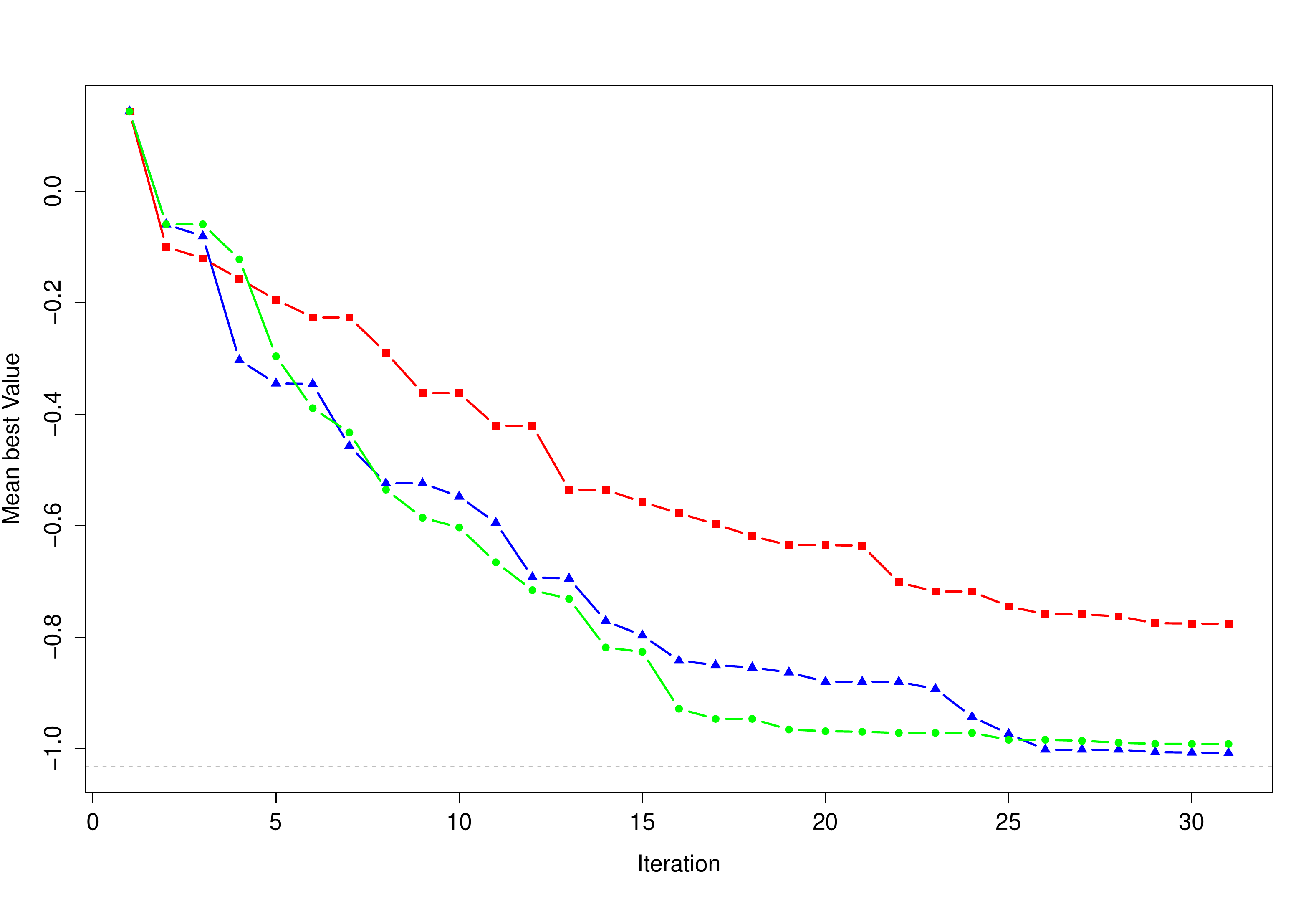}  \label{fig:opt2}} \par\medskip%  }
       \subfloat[][Ackley ]{ \includegraphics[width=0.45\textwidth,height = 5cm,natwidth=10.15in,natheight=8.61in]{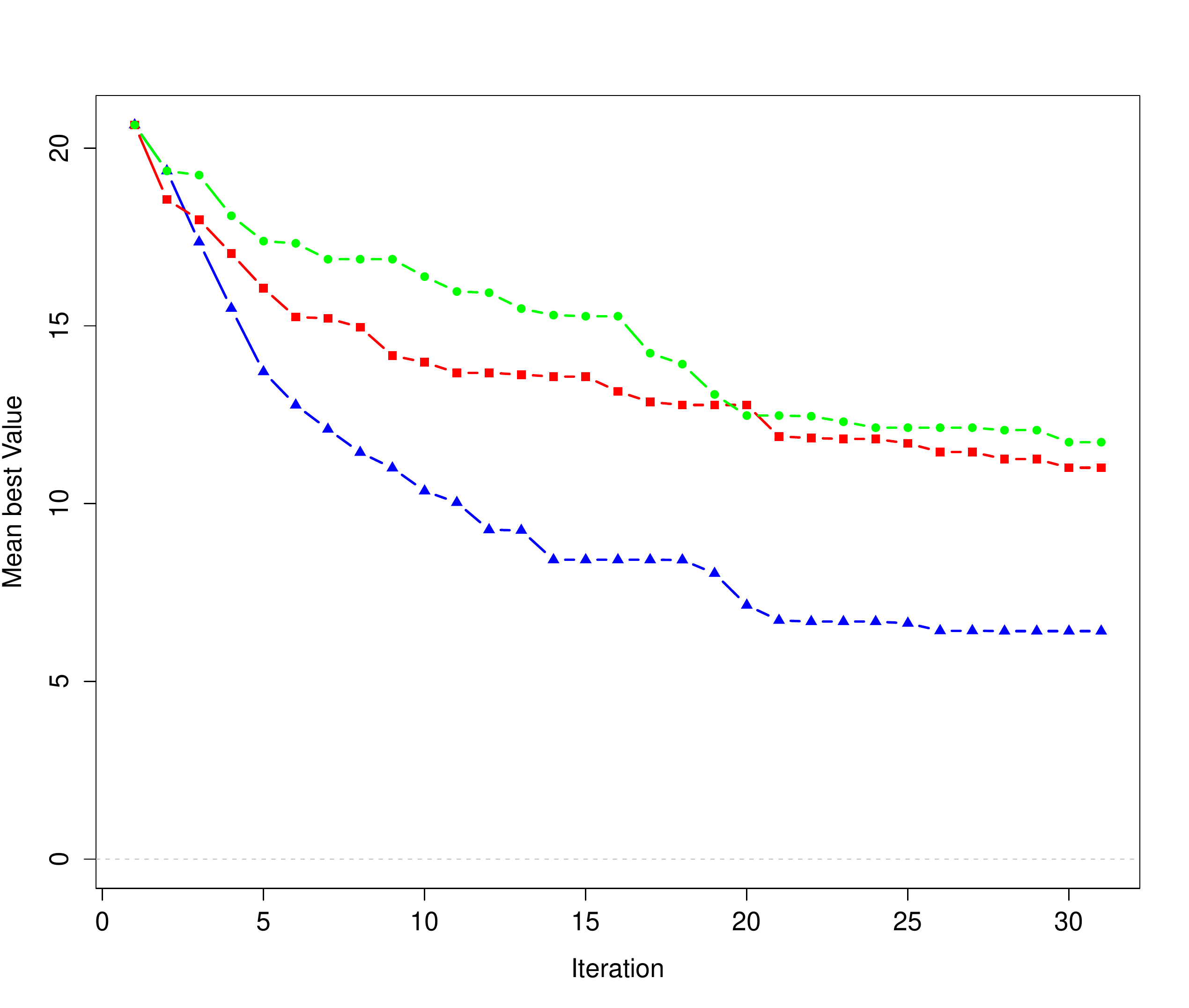} \label{fig:opt3}}%   \}
           \subfloat[][Hartmann6]{ \includegraphics[width=0.45\textwidth,height = 5cm,natwidth=19.35in,natheight=8.61in]{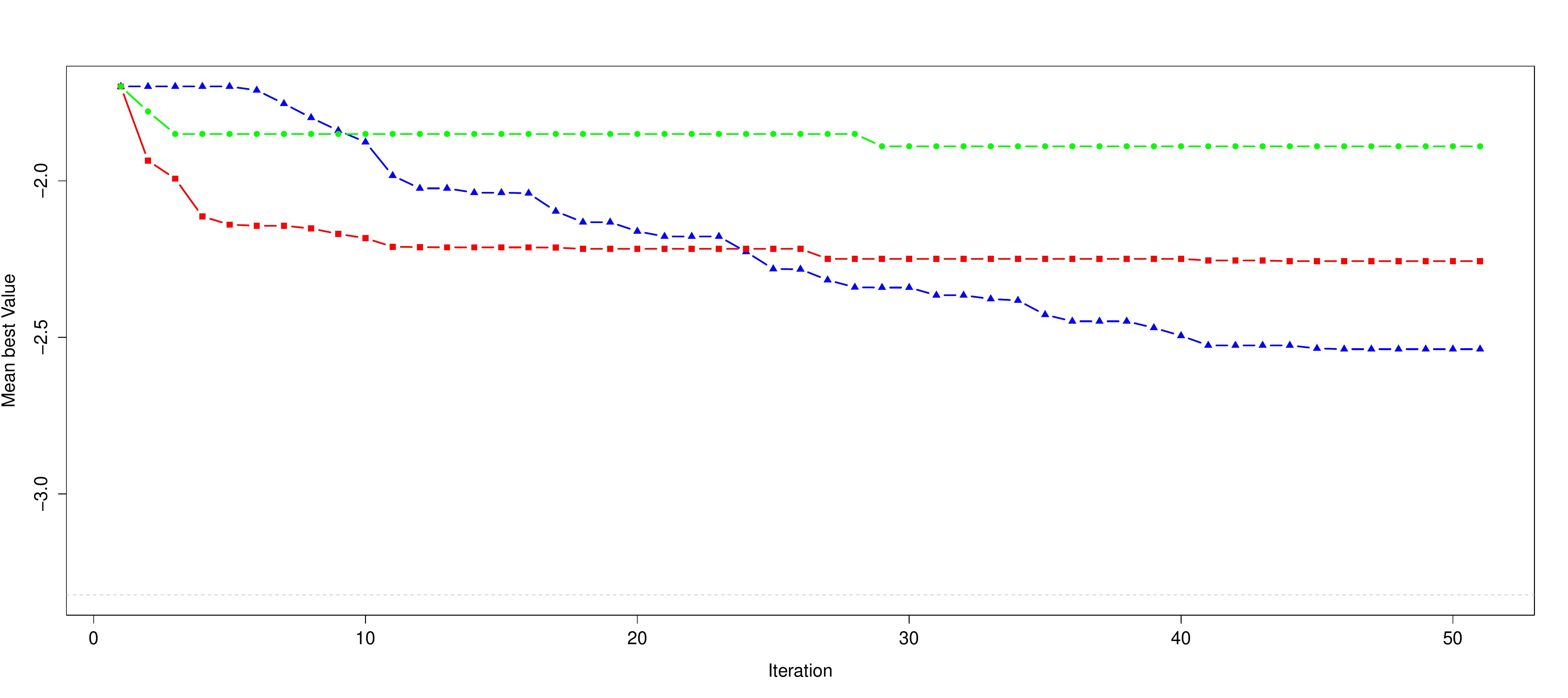}\label{fig:opt4}}%    \}
   \caption{ Comparison of three surrogate-based optimization strategies.
   Mean over  $N_{seed}$ of the best value   as a function of the number of iterations. UP-EGO with kriging in green,  UP-EGO with genetic aggregation in blue,  EGO in red and theoretical minimum in dashed gray.}  \label{fig:optmbv}
\end{figure}

For each function $f^{(i)}$, we launched each optimization process for $N_{max}^{(i)}$ iterations starting with $N_{seed} = 20$ different initial design of experiments of size  $n_0^{(i)}$  generated by an optimal space-filling sampling. The results are given using boxplots in Appendix \ref{sec:append}. We also display  the mean best value evolution in Figure \ref{fig:optmbv}.

The results shows that  the  UP-EGO  algorithms give  better results than the  EGO algorithm for Branin  and Camel functions. These cases illustrate the efficiency of the method. Moreover, for Ackley and Harmtann6 functions the best results are given by UP-EGO using the genetic aggregation.  Even if this is related to the nature of the surrogate model,  it underlines the efficient contribution of the  universality of UP-EGO.
Further,  let us focus on the boxplots of the last iterations of  Figures \ref{fig:bpopt1}  and   \ref{fig:bpopt4}  (Appendix \ref{sec:append}). It is important to notice that  UP-EGO results for Branin function   depend  slightly on  the initial design points.  On the other hand, let us focus on the  Hartmann function case.  The results of UP-EGO using the genetic aggregation  depend on the initial design points. In fact, more optimization iterations are required for a full convergence. However,  compared to the EGO algorithm, UP-EGO algorithm has good performances for both cases: 
 \begin{itemize}
 \item Full convergence 
 \item Limited-budget optimization.
 \end{itemize}
 
   \begin{figure}[ht]
 \centering
 \includegraphics[height=5cm,width=8cm,natwidth=122mm,natheight=106.36mm]{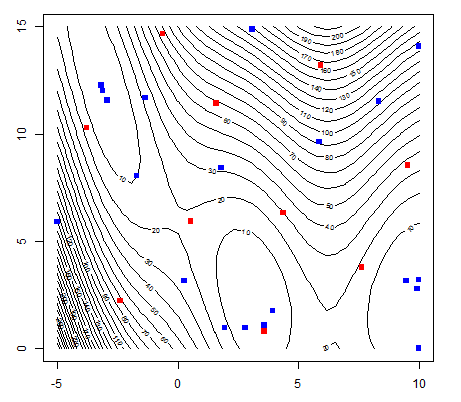}
 \caption{Example of sequence generated by by the  UP-EGO(kriging) algorithm on Branin function. Initial design points are in red, added points are in blue.}
 \label{fig:branincontour}
 \end{figure}
 
 Otherwise, the Branin function has multiple solutions. We are interested in checking whether the UP-EGO algorithm would focus on one local optimum or on the three possible  regions. We present in Figure \ref{fig:branincontour} the spatial distribution of the generated points  by the  UP-EGO(kriging) algorithm for the Branin function. We can notice that UP-EGO generated points around the three local minima.
 
\section{Fluid Simulation Application: Mixing Tank} \label{sec:mix}
 The problem addressed here concerns a static mixer where hot and cold fluid enter at variable velocities. 
The objective of this analysis is generally to find inlet velocities that minimize pressure loss from the cold inlet to the outlet and minimize the temperature spread at the outlet. In our study,  we are interested in a  better exploration of the design using an accurate cheap-to-evaluate surrogate model.
 
 \begin{figure}[ht]
 {\centering \includegraphics[height=4cm,width=.49\linewidth,natwidth=219.86mm,natheight=173.56mm]{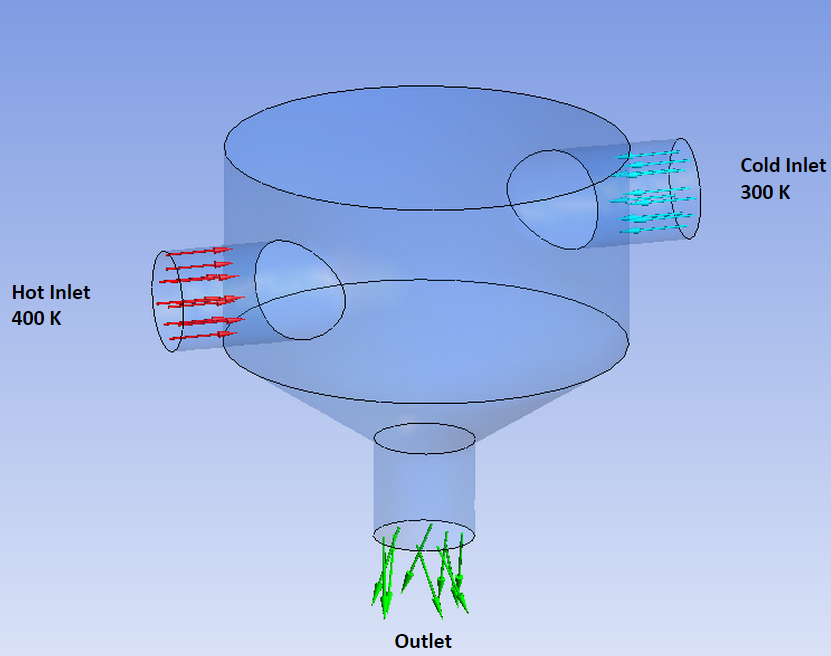} \includegraphics[height=4cm,width=.49\linewidth,natwidth=142.61mm,natheight=172mm]{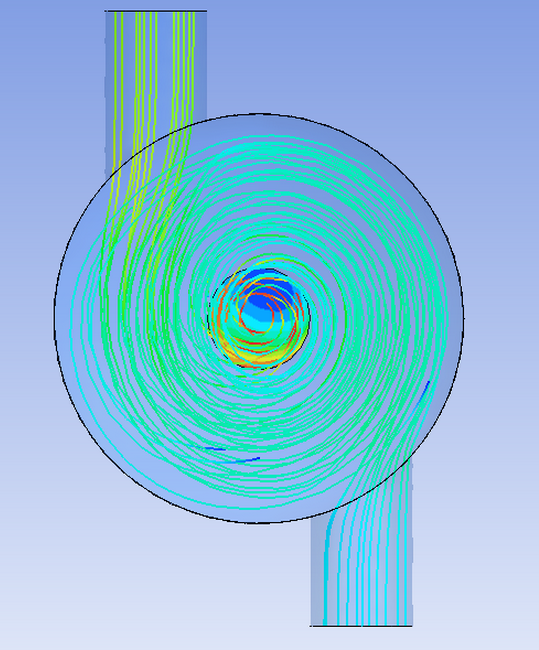}}

  \caption{Mixing tank} \label{fig:triangleMesh}
\end{figure}

 The simulations are computed within ANSYS Workbench environment and we used DesignXplorer  to perform surrogate-modeling. 
 We started the study using 9 design points generated by a central composite design. We produced also a set of $N_t =80$ test points $Z_t=(X_t =(x_1^{(t)}),\dots,\mathbf{x}_{N_t}^{(t)}),Y_t  =(y_1^{(t)}),\dots,y_{N_t}^{(t)}))$.  We launched   UP-SMART applied to the genetic aggregation response surface (GARS) in order to generate 10 suitable design points and a kriging-based refinement strategy. The  genetic aggregation response surface (GARS) developed by DesignXplorer  creates  a mixture of surrogate models including support vector machine regression, Gaussian process regression, moving Least Squares and polynomial regression. 
 We computed the root mean square error (Equation \eqref{eq:rmse}), the relative root mean square error (Equation \eqref{eq:rrmse}) and the relative average absolute error   (Equation \eqref{eq:raae}) before and after launching the refinement processes.
  \begin{align}
  RMSE_{Z^{(t)}}(\hat{s})  &= \frac{1}{N^t} \sum\limits_{i=1}^ {N_t}  (y_i^{(t)} - \hat{s}(\mathbf{x_i}^{(t)}))^2  \label{eq:rmse}\\
  RRMSE_{Z^{(t)}}(\hat{s})   &= \frac{1}{N^t}\sum\limits_{i=1}^{N_t} \left(\frac{ y_i^{(t)} - \hat{s}(\mathbf{x_i}^{(t)})}{y_i^{(t)}}\right)^2 \label{eq:rrmse}\\
  RAAE_{Z^{(t)}}(\hat{s})   &= \frac{1}{N^t}  \sum\limits_{i=1}^{N_t} \frac{ \mid y_i^{(t)} - \hat{s}(\mathbf{x_i}^{(t)}) \mid }{  \sigma_{Y}}\label{eq:raae}
  \end{align}
  We give  in Table \ref{tab:mixingtank}  the obtained  quality measures for the temperature spread output. In fact, the pressure loss is nearly linear and every method gives a good approximation.

\begin{table}[ht]
\begin{center}
\caption{ Quality measures of different response surfaces of static mixer simulations }
\label{tab:mixingtank} % Give a unique label
\begin{tabular}{p{0.25\textwidth}p{0.2\textwidth}p{0.15\textwidth}p{0.14\textwidth}} 
\hline%\noalign{\smallskip}
\hline
Surrogate model &  RRMSE &  RMSE & RAAE\\
%\noalign{\smallskip}
\hline%\noalign{\smallskip}
GARS Initial & 0.16 & 0.10 & 0.50\\
 
GARS Final & 0.10 & 0.07 &0.31\\
 
Kriging Initial  & 0.16 & 0.11 & 0.48\\

Kriging Final & 0.16 & 0.11 &0.50\\
%\noalign{\smallskip}
\hline
\end{tabular}
\end{center}
\end{table}

The results show that  UP-SMART gives a better approximation.  Here, it is used with a genetic aggregation of several response surface. Even  if the good quality may be due to the response surface itself, it highlights the fact that UP-SMART made the use of such surrogate model-based refinement strategy possible.
 
\section{Empirical Inversion} \label{sec:empinversion}
\subsection{Empirical inversion criteria  adaptation}
Inversion approaches consist in  the estimation of contour lines,  excursion sets or probability of failure. % of an expensive-to-evaluate function.
These techniques are specially used in  constrained  optimization  and  reliability  analysis.

Several iterative sampling strategies have been proposed to handle these problems. The empirical distribution $\mu_{n,\mathbf{x}}$ can be used for inversion problems. In fact, we can compute most of the well-known criteria such as the Bichon's criterion \cite{bichon2008efficient} or the Ranjan's criterion \cite{ranjan2008} using the UP distribution. In this section, we discuss   some of these  criteria: the targeted mean square error $TMSE$  \cite{picheny2010}, Bichon  \cite{bichon2008efficient} and the Ranjan criteria \cite{ranjan2008}. %as point-wise sampling criteria and the stepwise uncertainty reduction (SUR) strategy  \cite{bect2012}  as a  global sampling strategy.
The reader can refer to  Chevalier et al. \cite{ChevalierkrigInv} for an overview.

Let us consider the contour line estimation problem : let $T$ be a fixed threshold. We are interested in enhancing the surrogate model accuracy  in  $\{  \mathbf{x}   \in \mathbb{X}, s(\mathbf{x}) = T  \}$ and in its neighborhood.
\paragraph{Targeted MSE (TMSE)}
The targeted  Mean Square Error (TMSE) \cite{picheny2010} aims at decreasing the mean square error where the kriging prediction is close to T.
 
It is the  probability that the response lies inside the interval $ \big[T-\varepsilon, T+\varepsilon \big]$ where  the parameter $\varepsilon>0$  tunes the size of the  window  around the threshold $T$. High values make the criterion more exploratory while low values concentrate the evaluation around the contour line.
 
 We can  compute an estimation of the value of this criterion using the \textit{UP distribution} (Equation  \eqref{eq:timse2}).
\begin{equation} \label{eq:timse2}
\begin{aligned}
  TMSE_{T,n}(\mathbf{x})& = \sum\limits_{i=1}^n w_{i,n}(\mathbf{x})  1_{ \big[T-\varepsilon, T+\varepsilon \big]} \big(\hat{s}_{n,-i}(\mathbf{x})\big)\\ %  \hat{\sigma}_{n}(\mathbf{x})
 &=  \sum\limits_{i=1}^n w_{i,n}(\mathbf{x})  1_{ \big[-\varepsilon, \varepsilon \big]} \big(\hat{s}_{n,-i}(\mathbf{x}) - T\big)  % \hat{\sigma}_{n}(\mathbf{x})
  \end{aligned}
 \end{equation}
 Notice that the last criterion  takes into account neither the variability of the predictions at $\mathbf{x}$ nor the magnitude of the distance between the predictions and $T$.

\paragraph{Bichon criterion}

The expected feasibility defined in \cite{bichon2008efficient} 
%(Equation  \eqref{eq:bichonOr} ) 
aims at indicating  how well the true value of
the response is expected to be close to  the threshold $T$. 
 
The bounds are defined by  $\varepsilon_{\mathbf{x}}$ which  is proportional to the kriging  standard deviation $ \hat{\sigma}(\mathbf{x})$. Bichon proposes using $\varepsilon_{\mathbf{x}} = 2 \hat{\sigma}(\mathbf{x})$  \cite{bichon2008efficient}.
 
 This criterion can be extended to the case of the UP distribution. We define in  Equation  \eqref{eq:bichon} $EF_n$ the empirical Bichon's criterion where  $\varepsilon_{\mathbf{x}}$  is  proportional to the empirical standard deviation $  \hat{\sigma}^2_{n}(\mathbf{x})$ (Equation  \eqref{eq:var}). 
\begin{equation}
\begin{aligned}
EF_n(\mathbf{x}) 
& =  \sum\limits_{i=1}^n w_{i,n}(\mathbf{x})  (\varepsilon_{\mathbf{x}} - |T - \hat{s}_{n,-i}(\mathbf{x})|)  1_{ \big[-\varepsilon_{\mathbf{x}}, \varepsilon_{\mathbf{x}} \big]}(\hat{s}_{n,-i}(\mathbf{x}) - T )
\label{eq:bichon}
\end{aligned}
\end{equation}

\paragraph{Ranjan criterion}
 Ranjan et al. \cite{ranjan2008} proposed a criterion that quantifies  the improvement $I_{Ranjan}(\mathbf{x})$ defined in Equation  \eqref{eq:ranjanimp} 
\begin{equation}
\label{eq:ranjanimp}
I_{Ranjan}(\mathbf{x})= \left(\varepsilon^2_{\mathbf{x}} - (y(\mathbf{x})-T)^2 \right)   1_{ [-\varepsilon_{\mathbf{x}}, \varepsilon_{\mathbf{x}}]} (y(\mathbf{x}) - T) 
\end{equation}
where $\varepsilon_{\mathbf{x}} = \alpha  \hat{\sigma}(\mathbf{x})$,  and $\alpha>0$. $\varepsilon_{\mathbf{x}}$ defines the size of  the neighborhood around the contour $T$.

It is possible to compute the UP distribution-based  Ranjan's criterion (Equation  \eqref{eq:ranjanemp}). Note that   we set $\varepsilon_{\mathbf{x}} =\alpha    \hat{\sigma}^2_{n}(\mathbf{x})$.

\begin{equation}
\begin{aligned}
\mathbb{E} \Big[  I_{Ranjan}(\mathbf{x}) \Big] % EI_{Ranjan}(\mathbf{x}) &=\\
 &=  \sum\limits_{i=1}^n  w_{i,n}(\mathbf{x})  \Big(\varepsilon_{\mathbf{x}}^2 - (\hat{s}_{n,-i}(\mathbf{x})-T)^2\Big) 1_{ [-\varepsilon_{\mathbf{x}}, \varepsilon_{\mathbf{x}}]} (\hat{s}_{n,-i}(\mathbf{x}) - T)
\label{eq:ranjanemp}
\end{aligned}
\end{equation}

\subsection{Discussion}
The use of the pointwise criteria (Equations  \eqref{eq:timse2}, \eqref{eq:bichon}, \eqref{eq:ranjanemp}) might face problems when the region of interest is relatively small to the prediction jumps. In fact, as the cumulative distribution function of the UP distribution  is a step function, the probability of the prediction being inside an interval can vanish even if it is around the mean value. For instance  $\mu_{n,\mathbf{x}}\big(y(\mathbf{x}) \in [ T-\varepsilon, T+\varepsilon ]\big)$ can be zero. This is  one of the drawbacks of the empirical distribution.  Some regularization techniques  are possible to overcome this problem. 
For instance, the technique that consists in defining the region of interest by a Gaussian density $ \mathcal{N} (0,\sigma^2_\varepsilon)$ \cite{picheny2010}.  Let  $g_\varepsilon$ be this Gaussian probability distribution function.

The new $TMSE$ denoted $TMSE^{(2)}_{T,n}(\mathbf{x})$ criterion is then as in Equation  \eqref{eq:timse3}.
\begin{equation} \label{eq:timse3}
  TMSE^{(2)}_{T,n}(\mathbf{x}) =   \sum\limits_{i=1}^n w_{i,n}(\mathbf{x}) g_\varepsilon \big(\hat{s}_{n,-i}(\mathbf{x}) - T\big) % \hat{\sigma}_{n}(\mathbf{x})
 \end{equation} 
 
 The use of the Gaussian density to define the targeted region seems more relevant when using the UP local varaince. Similarly, we can apply the same method to the Ranjan's and Bichon's criteria. 
\section{Conclusion} \label{sec:conclusion}

To perform surrogate model-based sequential sampling, several relevant techniques require to quantify the prediction uncertainty associated to the model. Gaussian process regression provides directly this uncertainty quantification. This is the reason why Gaussian modeling is quite popular in sequential sampling.  
In this work, we defined a universal approach for uncertainty quantification that could be applied for any surrogate model. It is based on a weighted empirical probability measure supported by cross-validation sub-models predictions. 
 
Hence, one could use this distribution to compute most of the  classical sequential sampling criteria. As examples, we discussed sampling strategies for refinement, optimization and inversion.  
Further, we showed that, under some assumptions,  the optimum is adherent to the sequence of points generated by the optimization algorithm  UP-EGO.   Moreover, the optimization  and the refinement algorithms were successfully  implemented and tested both on  single and multiple surrogate models. We also discussed  the adaptation of some inversion criteria.
 The main drawback of  UP distribution  is that it is supported by a finite number of points. To avoid this, we propose to regularize this probability measure.  In a future work,  we will study and implement such regularization scheme and study the case of multi-objective constrained optimization.  
 
\section{Proofs} \label{sec:proofs}
We present in this section the proofs of Proposition  \ref{prop:kappanull},  Lemma \ref{prop:wu} and Theorem \ref{propconv}. 
Here, we use the  notations of Section \ref{sec:upegoconv}.

\begin{proof}[Proposition \ref{prop:kappanull}]
 Let $n > 1$, $\mathbf{Z_n}=(\mathbf{X_n}=(\mathbf{x_1},\dots,\mathbf{x_n})^\top,\mathbf{Y_n} = s(\mathbf{X_n})),$  and $\hat{s}$  a model that interpolates the data i.e  $\forall i \in 1,\dots,n$,  $ \hat{s}_{\mathbf{Z_n}} (\mathbf{x_i}) = s(\mathbf{x_i}) = y_i$. 

 First, we have  $\xi_n(\mathbf{x_i}) =  \delta  \underline{d}_{\mathbf{X_n}}(\mathbf{x_i})$. Since $\mathbf{x_i} \in \mathbf{X_n}$ then $\xi_n(\mathbf{x_i}) = 0$ . Further,
$ EEI_n(\mathbf{x_i})  = w_{i,n}(\mathbf{x_i}) \max(y_n^\star- \hat{s}_{n,-i}(\mathbf{x_i}),0 ) +  \sum\limits_{\substack{j=1\\ j \neq i}}^n w_{j,n}(\mathbf{x_i}) \max(y_n^\star -y_i,0 ) $. 
Notice that   $ w_{i,n}(\mathbf{x_i}) =0$ and   $ \max(y_n^\star -y_i,0 ) =0$ 

Then $EEI_n(\mathbf{x_i}) = 0$.   Finally, $\kappa_n(\mathbf{x_i}) = EEI_n(\mathbf{x_i}) + \xi_n(\mathbf{x_i}) = 0$. 
\end{proof}

\begin{proof} [Lemma \ref{prop:wu}]
Let us  note :  \begin{itemize}
\item  $\phi_\rho(\mathbf{x},\mathbf{x'}) = 1- e^{- \frac{d((\mathbf{x},\mathbf{x'}))^2}{\rho^2}}$. 
\item  $w_{i,n}(\mathbf{x}) = \frac{\phi_\rho(\mathbf{x},\mathbf{x_i})}{\sum\limits_{k=1}^n \phi_\rho(\mathbf{x},\mathbf{x_k})}$.
\end{itemize}

 Convex inequality gives   $ \forall a \in \mathbb{R}$,  $ 1 -e^{-a} < a$ then  $\phi_\rho(\mathbf{x},\mathbf{x_k}) \leq \frac{d((\mathbf{x},\mathbf{x_k}))^2}{\rho^2}$. %  
 Further, let $\mathbf{x_{k_1}},\mathbf{x_{k_2}}$ be two different design points of $ X_{n_0}$,  $ \forall \mathbf{x} \in \mathbb{X}$,   $\max\limits_{i \in \{1,2\}} \{ d(\mathbf{x} , \mathbf{x_{k_i}})\}\geq \frac{d(\mathbf{x_{k_1}},\mathbf{x_{k_2}})}{2}$ %or $d(\mathbf{x},\mathbf{x_{k_2}})\geq \frac{d(\mathbf{x'},\mathbf{x_{k_2}})}{2}$ s
 otherwise the triangular inequality would be violated. Consequently, 
 
 $\forall n> n_0$, $\sum\limits_{k=1}^n \phi_\rho(\mathbf{x},\mathbf{x_k})\geq \phi_\rho(\mathbf{x},\mathbf{x_{k_1}}) + 
 \phi_\rho(\mathbf{x},\mathbf{x_{k_2}}) \geq \phi_{2\rho}(\mathbf{x_{k_1}},\mathbf{x_{k_2}})>0$
 
  $\forall n> n_0$, $\forall \mathbf{x} \in \mathbb{X}$: 
 $w_{i,n}(\mathbf{x})  = \frac{\phi_{i,n}(\mathbf{x})}{\sum\limits_{k=1}^n \phi_{k,n}(\mathbf{x})}  \leq   \frac{\phi_{i,n}(\mathbf{x})}{ \phi_{2\rho}(\mathbf{x_{k_1}},\mathbf{x_{k_2}})} \leq   \frac{d((\mathbf{x},\mathbf{x_i}))^2}{\rho^2 \phi_{2\rho}(\mathbf{x_{k_1}},\mathbf{x_{k_2}})} $
 
 Considering $\theta = \frac{1}{\rho^2 \phi_{2\rho}(\mathbf{x_{k_1}},\mathbf{x_{k_2}}) }$ ends the proof.
 \end{proof}

\begin{proof}[Theorem \ref{propconv}]
$\mathbb{X}$ is compact so $S$  has a convergent  sub-sequence  in $\mathbb{X}^{\mathbb{N}}$   (Bolzano-Weierstrass theorem  ). Let $(x_{\psi (n)})$  denote  that sub-sequence and  $\mathbf{x_{\infty}} \in \mathbb{X}$ its limit.  
 %Notice that $(\kappa_n)$   is equicontinuous.  
%and then since $\mathbf{x_{\infty}}$ is adherent to the sequence $S$, $\max\limits_{\mathbf{x} \in \mathbb{X}} \{ \kappa_{\psi (n)-1}(\mathbf{x}) \} =  \kappa_{\psi (n)-1}(\mathbf{x}_{\psi (n)}) $ $\to 0$ when $n \to \infty$. This ends the proof since $ \mathbf{x^\star} \in\mathbb{X} $.
We can assume by considering a sub-sequence of $\psi$ and using the continuity of  the surrogate model $\hat{s}$ that: 
\begin{itemize}
\item $d(\mathbf{x_\infty},\mathbf{x}_{\psi (n)}) \leq \frac{1}{n}$ for all $n>0$
\item $\exists \nu_n \geq  d(\mathbf{x_{\infty}},\mathbf{x}_{\psi (n)})$ such that $\forall \mathbf{x'} \in \mathbb{X} $, $d(\mathbf{x'},\mathbf{x_\infty}) \leq \nu_n  \implies \left| \hat{s}_{m,-i}(\mathbf{x_{\infty}})  -  \hat{s}_{m,-i}(\mathbf{x'})\right| \leq  \frac{1}{n},$ $ \forall i \in 1,\dots,m$, where $m> n_0$.
\end{itemize}

For all $k>1$, we note $v_k = \psi(k+1)-1$, the step at which UP-EGO algorithm selects the point $\mathbf{x}_{\psi(k+1)}$.
So, $ \kappa_{v_k}(\mathbf{x}_{\psi (k+1)})  =\max\limits_{\mathbf{x} \in \mathbb{X}} \{ \kappa_{v_k}(\mathbf{x}) \}$. 

Notice first that  for all $n>0$, $\mathbf{x}_{\psi(n)}, \mathbf{x}_{\psi(n+1)} \in \mathcal{B}(\mathbf{x_{\infty}}, \frac{1}{n})$ where $\mathcal{B}(\mathbf{x_{\infty}}, \frac{1}{n})$ is  the closed ball of center $\mathbf{x_{\infty}}$ and radius $\frac{1}{n}$. So: 
\begin{equation}
  \tag{i}
    \xi_{v_n}(\mathbf{x}_{\psi(n+1)}) = \delta \underline{d}_{X_{v_n}}(\mathbf{x}_{\psi(n+1)})   \leq  \delta d(\mathbf{x_{\psi(n)}},\mathbf{x}_{\psi(n+1)}) \leq \frac{2\delta }{n} 
  \label{eqn:1}
\end{equation}
According to Lemma \ref{prop:wu}, $ w_{\psi(n), v_n}\leq \theta \left(d(\mathbf{x}_{\psi(n+1)},\mathbf{x}_{\psi(n)})\right)^2$ so $ w_{\psi(n), v_n} \leq \frac{4\theta}{n^2}$.   
Consequently:
\begin{equation}
  \tag{ii}
 w_{\psi(n), v_n}(\mathbf{x}_{\psi(n+1)}) \max(y_{v_n}^\star - \hat{s}_{v_n,-\psi(n)}(\mathbf{x}_{\psi(n+1)}),0) \leq  \frac{4\theta (U-L)}{n^2} 
 \label{eqn:2}
\end{equation}

Further, $\forall i \in 1,\dots,v_n$, $i \neq  \psi(n)$,  $\hat{s}_{v_n,-i}(\mathbf{x}_{ \psi(n)})  = y_{\psi(n)}$ since the surrogate model is an interpolating one. hence $ \hat{s}_{v_n,-i}(\mathbf{x}_{ \psi(n)})  \geq y_{v_n}^\star $ and so  
 $ \max(y_{v_n}^\star - \hat{s}_{v_n,-i},0)   \leq   \max(\hat{s}_{v_n,-i}(\mathbf{x}_{ \psi(n)}) - \hat{s}_{v_n,-i}(\mathbf{x}_{ \psi(n+1)}),0)  \leq  \left| \hat{s}_{v_n,-i}(\mathbf{x}_{ \psi(n)}) - \hat{s}_{v_n,-i}(\mathbf{x}_{ \psi(n+1)})  \right|$. Triangular inequality gives: $\max(y_{v_n}^\star - \hat{s}_{v_n,-i},0)  \leq \left| \hat{s}_{v_n,-i}(\mathbf{x}_{ \psi(n)}) - \hat{s}_{v_n,-i}(\mathbf{x}_{\infty})  \right|+ \left| \hat{s}_{v_n,-i}(\mathbf{x}_{\infty}) - \hat{s}_{v_n,-i}(\mathbf{x}_{ \psi(n+1)})\right| $ and finally:
 \begin{equation}
  \tag{iii}
 \max(y_{v_n}^\star - \hat{s}_{v_n,-i},0)    \leq  \frac{2}{n} 
  \label{eqn:3}
\end{equation}
We have:
\begin{equation*}
\begin{aligned}
 \left| \kappa_{v_n}(\mathbf{x}_{ \psi(n+1)}) \right| %&= \kappa_{u_k}(\chi_{k+1}) - \kappa_{u_k}(\chi_k)\\ 
&=  \xi_{v_n}(\mathbf{x}_{ \psi(n+1)})  + \sum\limits_{i=1}^{v_n} w_{i,v_n}(\mathbf{x}_{ \psi(n+1)}) \max(y_{v_n}^\star -   \hat{s}_{v_n,-i}(\mathbf{x}_{ \psi(n+1)}),0 )\\ % - w_{i,k}(\chi_k) \max(y_{u_k}^\star -   \hat{s}_{u_k,-i}(\chi_k),0 )\\
& = \xi_{v_n}(\mathbf{x}_{ \psi(n+1)})  + w_{ \psi(n),v_n}(\mathbf{x}_{ \psi(n+1)}) \max(y_{v_n}^\star -   \hat{s}_{v_n,- \psi(n)}(\mathbf{x}_{ \psi(n+1)}),0 )\\
 &+ \sum\limits_{\substack{i=1\\ i \neq  \psi(n)}}^{v_n} w_{i,v_n}(\mathbf{x}_{ \psi(n+1)}) \max(y_{v_n}^\star  -   \hat{s}_{v_n,-i}(\mathbf{x}_{ \psi(n+1)}),0 )\\   % - w_{i,k}(\chi_k) \max(y_{u_k}^\star -   \hat{s}_{u_k,-i}(\chi_k),0 )\\
 &\leq \frac{2\delta}{n}  + \frac{4\theta (U-L)}{n^2} +  \frac{2}{n}\\
% &+ \sum\limits_{\substack{i=1\\ i \neq u_k}}^{u_k} w_{i,v_k}(\chi_{k+1}) \max(y_{v_k}^\star  -   \hat{s}_{v_k,-i}(\chi_{k+1}),0 )   
\end{aligned}
\end{equation*}
Considering \eqref{eqn:1},\eqref{eqn:2} and \eqref{eqn:3} :
\begin{equation*}
 \left| \kappa_{v_n}(\mathbf{x}_{ \psi(n+1)}) \right|  \leq \frac{2\delta}{n}  + \frac{4\theta (U-L)}{n^2} +  \frac{2}{n}\\
\end{equation*}
 Notice that:\\
$\kappa_{v_n}(\mathbf{x}_{\psi(n+1)})  =\max\limits_{\mathbf{x} \in \mathbb{X}} \{ \kappa_{v_n}(\mathbf{x}) \}$
and  $ \delta  \underline{d}_{\mathbf{S}_{v_n}}(\mathbf{x^\star}) = \xi_{v_n}(\mathbf{x^\star})$ $ \leq \kappa_{v_n}(\mathbf{x^\star})  \leq \kappa_{v_n}(\mathbf{x}_{\psi(n)}) $. Since $ \lim\limits_{n\to \infty} \left| \kappa_{v_n}(\mathbf{x}_{\psi(n+1)}) \right| = 0$ so $ \lim\limits_{n\to \infty} \underline{d}_{\mathbf{S}_{v_n}}(\mathbf{x^\star}) \to 0$.
\end{proof}

 \section{Acknowledgment}
 We gratefully acknowledge the French National Association for Research and Technology (ANRT, CIFRE grant number 2015/1349).
 
 \appendix\section*{Appendix: Optimization test results} \label{sec:append} In this section, we  use boxplots to display the  evolution of the best value of the optimization test bench. For each iteration, we display: left: EGO in red., middle UP-EGO using genetic aggregation in blue, right: UP-EGO using kriging  in green. 

\begin{figure}[ht]
     \centering
    \includegraphics[width=\textwidth,height = 4.5cm,natwidth=11.75in,natheight=8.61in]{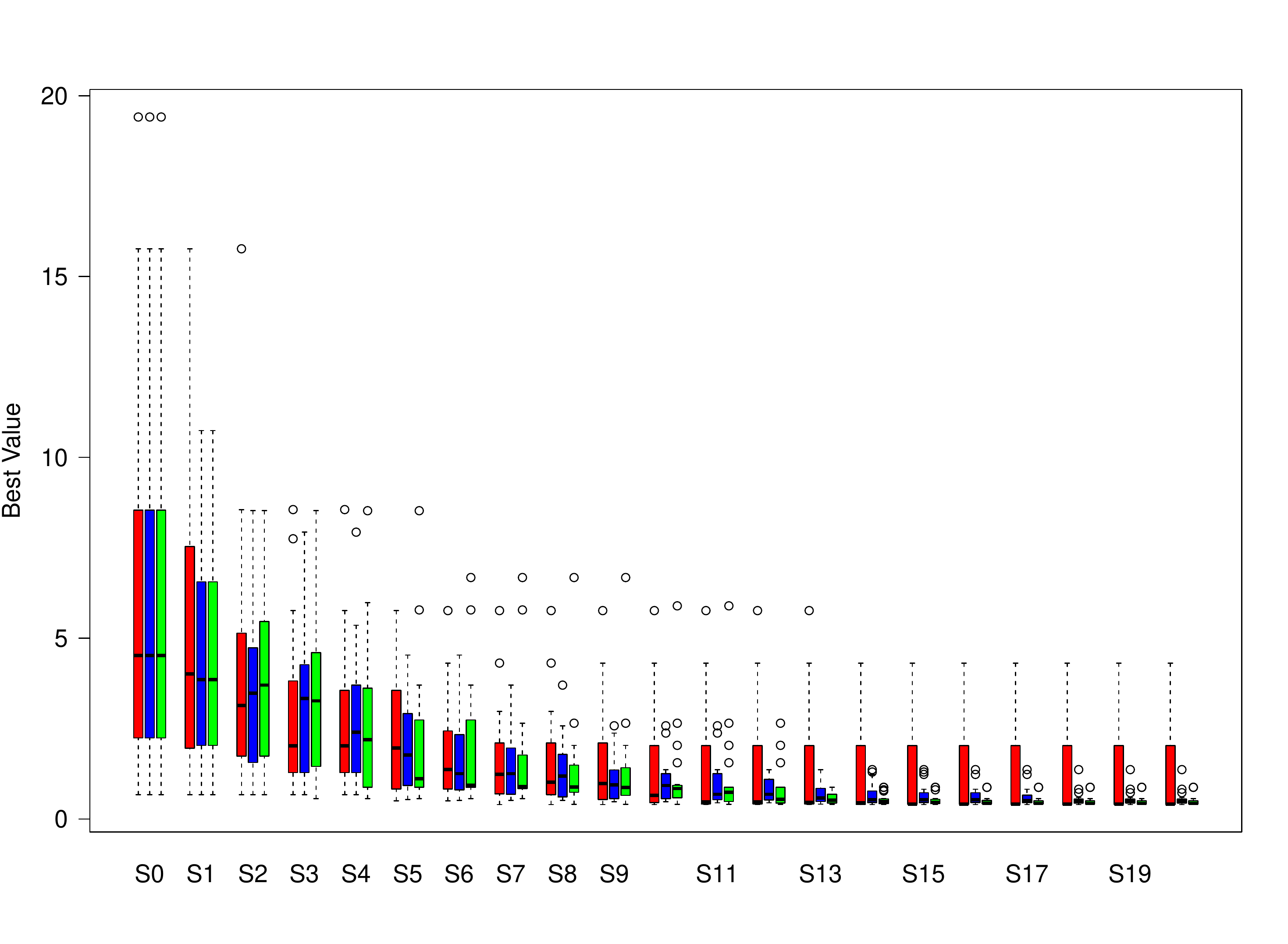}
 \caption{Branin: Box plots convergence}
    \label{fig:bpopt1}
 \end{figure}
 
\begin{figure}[ht]
    \centering
    \includegraphics[height=5cm,width=\textwidth,natwidth=18.21in,natheight=8.61in]{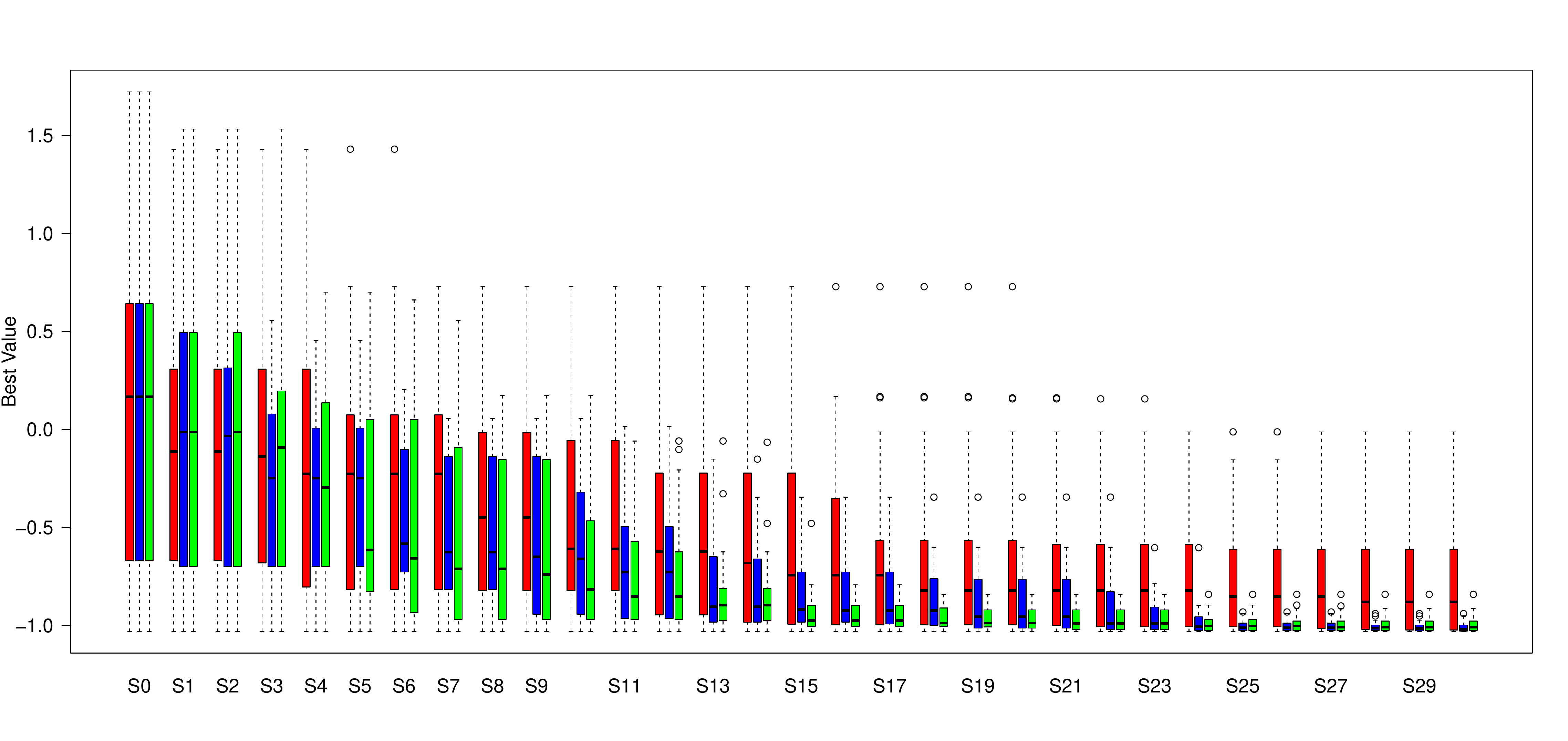}
 \caption{Six-hump camel: Box plots convergence}
    \label{fig:bpopt2}
\end{figure}
\begin{figure}[ht]
  \centering
    \centering\includegraphics[height=5cm,width=\textwidth,natwidth=18.36in,natheight=8.61in]{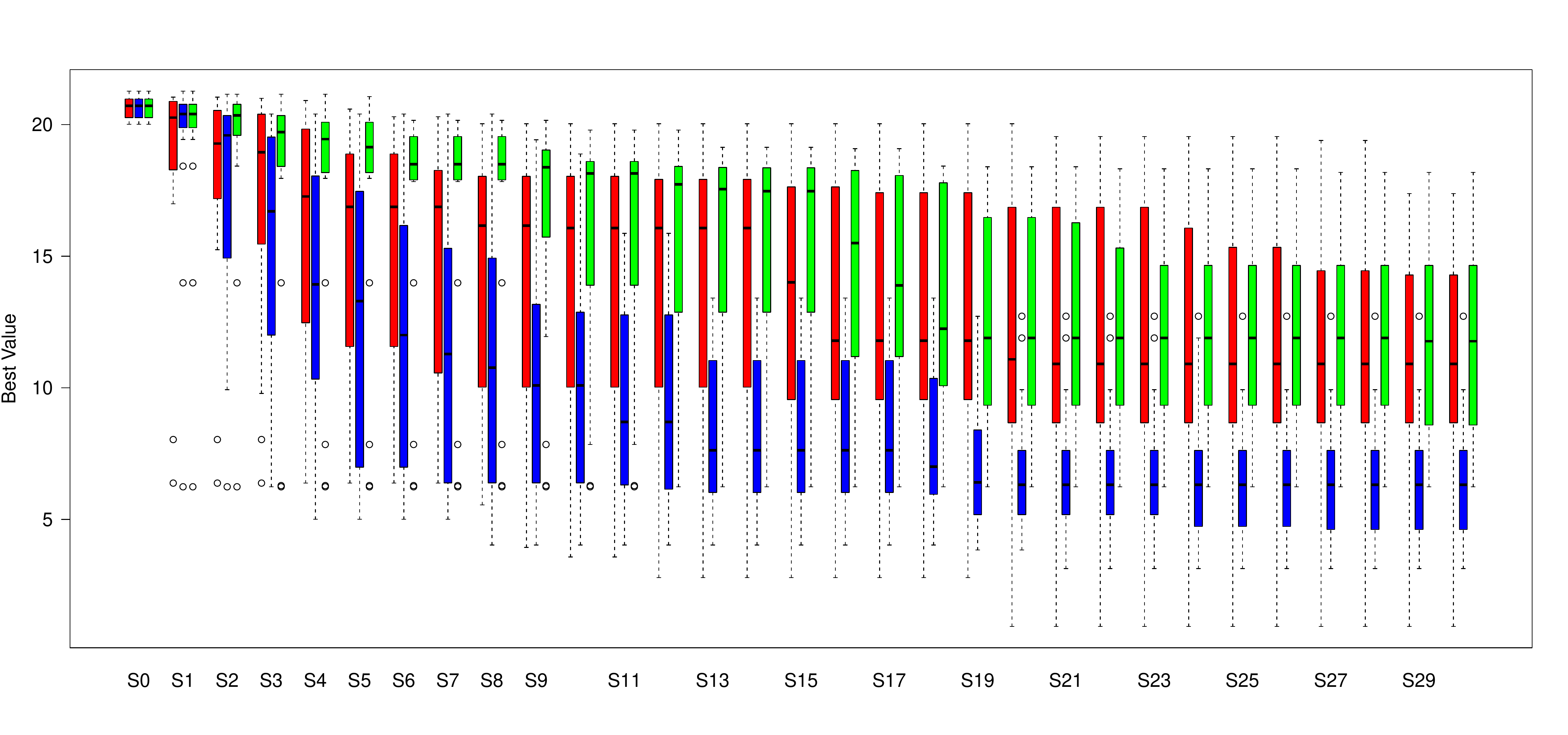}
 \caption{Ackley: Box plots convergence}
    \label{fig:bpopt3}
\end{figure}
\begin{figure}[!ht]
  \centering
    \centering\includegraphics[height=5.5cm,width=\linewidth,natwidth=19.35in,natheight=8.61in]{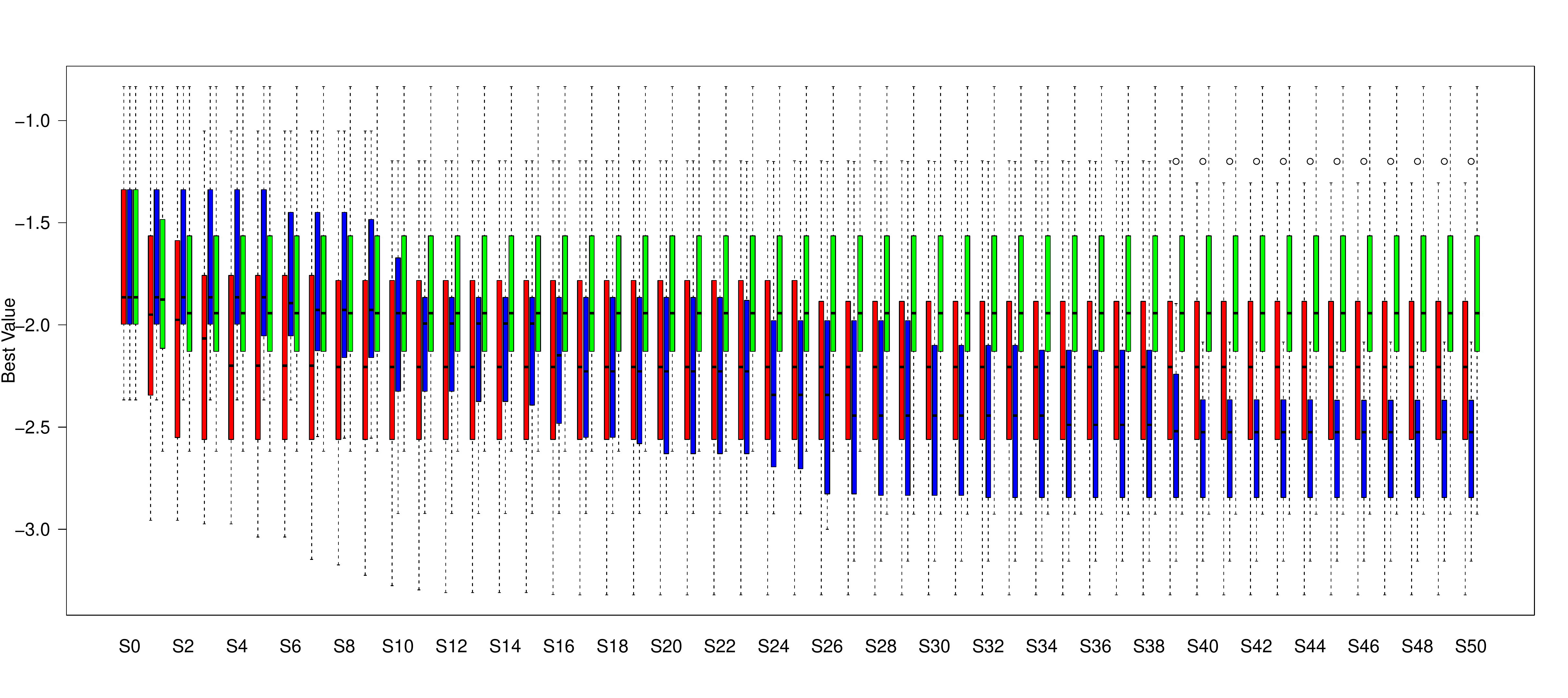}
 \caption{Hartmann6: Box plots convergence}
    \label{fig:bpopt4}
\end{figure}

%\bibliographystyle{siam.bst}
%\bibliography{biblio}{}

\end{document}